\documentclass[notitlepage,nofootinbib,aps,prd,a4paper,11pt,amsfonts,amssymb,amsmath]{revtex4-1}
\usepackage{bigints}

\begin{document}
\title{Parameterized post-Newtonian limit of Horndeski's gravity theory}

\author{Manuel Hohmann}
\email{manuel.hohmann@ut.ee}
\affiliation{Teoreetilise F\"u\"usika Labor, F\"u\"usika Instituut, Tartu \"Ulikool, Ravila 14c, 50411 Tartu, Estonia}

\begin{abstract}
We discuss the parameterized post-Newtonian (PPN) limit of Horndeski's theory of gravity, also known under the name generalized G-inflation or $\text{G}^2$-inflation, which is the most general scalar-tensor theory of gravity with at most second order field equations in four dimensions. We derive conditions on the action for the validity of the post-Newtonian limit. For the most general class of theories consistent with these conditions we calculate the PPN parameters \(\gamma(r)\) and \(\beta(r)\), which in general depend on the interaction distance \(r\) between the gravitating mass and the test mass. For a more restricted class of theories, in which the scalar field is massless, we calculate the full set of PPN parameters. It turns out that in this restricted case all parameters are constants and that the only parameters potentially deviating from observations are \(\gamma\) and \(\beta\). We finally apply our results to a number of example theories, including galileons and different models of Higgs inflation.
\end{abstract}
\maketitle

\section{Motivation}\label{sec:motivation}
The most striking observations in modern cosmology are the accelerating expansion of the universe~\cite{Riess:1998cb,Perlmutter:1998np,Ade:2015xua,Ade:2015rim}, whose cause has been named dark energy, and the homogeneity of the microwave background~\cite{Ade:2015bva,Ade:2015lrj}, which is conventionally attributed to an inflationary expansion of the very early universe~\cite{Guth:1980zm,Linde:1981mu}. However, both dark energy and inflation are yet unexplained phenomena. A large and important class of theories aiming to explain these phenomena is based on the introduction of a scalar field mediating gravity in addition to the usual metric degrees of freedom~\cite{Thiry:1948,Jordan:1959eg,Brans:1961sx,Bergmann:1968ve,Wagoner:1970vr}. Various theories belonging to this class have been successfully applied to cosmology~\cite{Fujii:2003pa,Faraoni:2004pi}.

An important feature of general relativity, which one wishes to retain also in scalar-tensor theories of gravity, is the fact that its gravitational field equations contain at most second order derivatives of the dynamical fields. This restriction is imposed since higher derivative theories will in general lead to instabilities and ghosts~\cite{Ostrogradski:1850}. One may therefore ask which is the most general class of scalar-tensor theories whose field equations are of at most second order. This question was answered already by Horndeski~\cite{Horndeski:1974wa}, although his work did not receive much attention until recently and was re-derived in a different, but equivalent formulation~\cite{Deffayet:2011gz,Kobayashi:2011nu,Gao:2011mz,Charmousis:2011bf}. Since then, many particular theories belonging to this class have been studied, in particular as models of inflation and dark energy~\cite{Joyce:2014kja}.

However interesting its performance in cosmology, a viable gravitational theory must also pass the tests on local scales, e.g., give a good account of the motions in our solar system. A natural framework for such a check is the parameterized post-Newtonian (PPN) formalism~\cite{Will:1993ns,Will:2014xja}. It characterizes gravity theories by a set of ten parameters, which have been measured with high precision in various solar system experiments~\cite{Fomalont:2009zg,Bertotti:2003rm,Hofmann:2010,Lambert:2011,Fienga:2011qh,Pitjeva:2013xxa,Verma:2013ata,Fienga:2014,Fienga:2014bvy}. Through the availability of this high precision data, the PPN formalism has become an important testbed for the viability of gravity theories.

This work extends and generalizes an earlier result on the PPN parameters \(\gamma\) and \(\beta\) for a class of scalar-tensor theories of gravity with a general potential in the Jordan frame~\cite{Hohmann:2013rba}, which have also been calculated in the Einstein frame taking into account screening effects~\cite{Scharer:2014kya} and in terms of invariants under conformal transformations and scalar field redefinitions~\cite{Jarv:2014hma}. Also the solar system physics of other theories belonging to the Horndeski class of gravity theories have already been studied and it has been argued that the ``fifth force'' mediated by the scalar degree of freedom should be suppressed in order to reproduce the observed general relativity limit. For this purpose, several screening mechanisms have been studied~\cite{Joyce:2014kja}, such as the chameleon~\cite{Khoury:2003rn,Khoury:2003aq,Gannouji:2010fc}, symmetron~\cite{Hinterbichler:2010es,Hinterbichler:2011ca} or Vainshtein~\cite{Vainshtein:1972sx,Boulware:1973my,Deffayet:2001uk} mechanisms, and it has been shown that these mechanisms can achieve consistency of the theory with solar system observations~\cite{Tsujikawa:2008uc,Kimura:2011dc,Kase:2013uja}.

In this work we complement these studies by an analysis of theories in which screening mechanisms do not play a significant role, so that the standard PPN formalism can be applied. We explicitly calculate the PPN parameters \(\gamma\) and \(\beta\) for a general class of Horndeski theories, and the full set of PPN parameters for a more restricted class with a massless scalar field, in order to show that also in this case consistency with solar system observations can be achieved, without employing any screening mechanisms.

The outline of this article is as follows. In section~\ref{sec:action} we display the action and discuss the structure of the field equations. In section~\ref{sec:pertexp} we expand these field equations in a weak field limit around a Minkowski background. The post-Newtonian limit of this expansion is discussed in section~\ref{sec:ppn}. The post-Newtonian gravitational field equations are then solved for a static point mass source in section~\ref{sec:sphersym}, which yields the PPN parameters \(\gamma\) and \(\beta\). The full set of PPN parameters is obtained in the case of a massless scalar field in section~\ref{sec:mlfullppn}. In section~\ref{sec:obs} we compare this general result to current observations of the PPN parameters. We apply our findings to a few example theories in section~\ref{sec:examples}, and end with a conclusion in section~\ref{sec:conclusion}.

\section{Action and field equations}\label{sec:action}
In this section we provide a brief overview of the structure of the action and the field equations of Horndeski's gravity theory. The starting point of our derivation is the action, which takes the form~\cite{Kobayashi:2011nu}
\begin{equation}\label{eqn:action}
S = \sum_{i = 2}^{5}\int d^4x\sqrt{-g}\mathcal{L}_i[g_{\mu\nu},\phi] + S_m[g_{\mu\nu},\chi_m]\,.
\end{equation}
Here \(S_m\) denotes the matter action and \(\chi_m\) collectively all matter fields. The gravitational part of the action, which depends on the metric \(g_{\mu\nu}\) and a single scalar field \(\phi\), is given as an integral over the four-dimensional spacetime manifold, where the Lagrangian is composed of the terms
\begin{gather}
\mathcal{L}_2 = K(\phi,X)\,,\quad
\mathcal{L}_3 = -G_3(\phi,X)\square\phi\,,\quad
\mathcal{L}_4 = G_4(\phi,X)R + G_{4X}(\phi,X)\left[(\square\phi)^2 - (\nabla_{\mu}\nabla_{\nu}\phi)^2\right]\,,\nonumber\\
\mathcal{L}_5 = G_5(\phi,X)G_{\mu\nu}\nabla^{\mu}\nabla^{\nu}\phi - \frac{1}{6}G_{5X}(\phi,X)\left[(\square\phi)^3 - 3(\square\phi)(\nabla_{\mu}\nabla_{\nu}\phi)^2 + 2(\nabla_{\mu}\nabla_{\nu}\phi)^3\right]\,.
\end{gather}
Here we introduced the notation
\begin{gather}
\square = g^{\mu\nu}\nabla_{\mu}\nabla_{\nu}\,,\quad
(\nabla_{\mu}\nabla_{\nu}\phi)^2 = \nabla_{\mu}\nabla_{\nu}\phi\nabla^{\mu}\nabla^{\nu}\phi\,,\nonumber\\
(\nabla_{\mu}\nabla_{\nu}\phi)^3 = \nabla_{\mu}\nabla_{\nu}\phi\nabla^{\nu}\nabla^{\lambda}\phi\nabla_{\lambda}\nabla^{\mu}\phi\,,\quad
X = -\frac{1}{2}\nabla^{\mu}\phi\nabla_{\mu}\phi
\end{gather}
for the d'Alembert operator \(\square\) and derivatives of the scalar field, and indices are raised and lowered with the metric \(g_{\mu\nu}\). The functions \(K, G_3, G_4, G_5\) are free functions of the scalar field \(\phi\) and its kinetic term \(X\). Each choice of these functions determines a distinct gravity theory. We denote derivatives of these functions by a subscript, e.g., \(G_{4X} = \partial G_4/\partial X\).

The gravitational field equations are derived from the action~\eqref{eqn:action} by variation with respect to the metric and the scalar field. It follows from the structure of the action that the field equations take the general form
\begin{equation}\label{eqn:fieldeqns}
\sum_{i = 2}^{5}\mathcal{G}^i_{\mu\nu} = \frac{1}{2}T_{\mu\nu}\,, \quad \sum_{i = 2}^{5}\nabla^{\mu}J^i_{\mu} = \sum_{i = 2}^{5}P^i_{\phi}\,,
\end{equation}
where \(T_{\mu\nu}\) is the energy-momentum tensor of the matter fields \(\chi_m\). The terms \(\mathcal{G}^i_{\mu\nu}\), \(J^i_{\mu}\) and \(P^i_{\phi}\) are obtained from the variation of the different Lagrangians in the gravitational part of the action. Their full form is rather lengthy and listed in the appendix of~\cite{Kobayashi:2011nu}. However, for practical purposes it turns out to be easier to replace the metric field equation with its trace-reversed analogue
\begin{equation}\label{eqn:trfieldeqns}
\sum_{i = 2}^{5}\mathcal{R}^i_{\mu\nu} = \frac{1}{2}\bar{T}_{\mu\nu} = \frac{1}{2}\left(T_{\mu\nu} - \frac{1}{2}g_{\mu\nu}T\right)\,,
\end{equation}
where the trace-reversed metric terms \(\mathcal{R}^i_{\mu\nu}\) are given by
\begin{equation}\label{eqn:trcurv}
\mathcal{R}^i_{\mu\nu} = \mathcal{G}^i_{\mu\nu} - \frac{1}{2}g_{\mu\nu}g^{\rho\sigma}\mathcal{G}^i_{\rho\sigma}\,.
\end{equation}
These are the field equations we will be working with in this article. For the purpose of calculating their post-Newtonian limit, we first need to bring them into a more manageable form using a perturbative expansion around a fixed background solution. This will be done in the next section.

\section{Perturbative expansion}\label{sec:pertexp}
In order to calculate the parameterized post-Newtonian limit of Horndeski's gravity theory we will need a perturbative expansion of the field equations, which we displayed in the preceding section, around a fixed background solution. This background solution will be given by a flat Minkowski metric \(\eta_{\mu\nu}\) and a constant cosmological background value \(\Phi\) of the scalar field, so that the perturbative expansion assumes the form
\begin{equation}
g_{\mu\nu} = \eta_{\mu\nu} + h_{\mu\nu}\,, \quad \phi = \Phi + \psi\,, \quad X = -\frac{1}{2}\nabla^{\mu}\psi\nabla_{\mu}\psi\,.
\end{equation}
Besides assuming that the background is homogeneous and isotropic, we thus also assume that it is stationary, i.e., constant in time. The physical reasoning behind this assumption is that we particularly consider the situation at or close to a fixed point of the background evolution of the scalar field, so that we can neglect any effects from a dynamical background. For the post-Newtonian limit it will be necessary to expand the terms \(\mathcal{G}^i_{\mu\nu}\) (and thus also \(\mathcal{R}^i_{\mu\nu}\)), \(J^i_{\mu}\) and \(P^i_{\phi}\) up to the quadratic order in the perturbations \(h_{\mu\nu}\) and \(\psi\) around this background. This will be done in this section.

Recall that the action of Horndeski's gravity theory, and thus also the field equations, depends on the choice of four free functions \(K, G_3, G_4, G_5\), which depend on the scalar field \(\phi\) and its kinetic term \(X\). Their Taylor expansion around the cosmological background value \(\Phi\) takes the form
\begin{equation}\label{eqn:taylorseries}
K(\phi,X) = \sum_{m,n = 0}^{\infty}K_{(m,n)}\psi^mX^n\,,
\end{equation}
where the coefficients \(K_{(m,n)}\) are given by
\begin{equation}\label{eqn:taylorcoeff}
K_{(m,n)} = \frac{1}{m!n!}\left.\frac{\partial^{m + n}}{\partial\phi^m\partial X^n}K(\phi,X)\right|_{\phi = \Phi, X = 0}\,,
\end{equation}
and similarly for the remaining functions \(G_3, G_4, G_5\). Each term \(K_{(m,n)}\psi^mX^n\) is of the order \(\mathcal{O}(\psi^{m+2n})\). Using these expansions, the terms constituting the field equations listed in~\cite{Kobayashi:2011nu} take the form
\begin{subequations}\label{eqn:gexpansion}
\begin{align}
\mathcal{G}^2_{\mu\nu} &= -\frac{1}{2}K_{(0,0)}\eta_{\mu\nu} - \frac{1}{2}K_{(0,0)}h_{\mu\nu} - \frac{1}{2}K_{(1,0)}\eta_{\mu\nu}\psi - \frac{1}{2}K_{(1,0)}h_{\mu\nu}\psi - \frac{1}{2}K_{(2,0)}\eta_{\mu\nu}\psi^2\nonumber\\
&\phantom{=}+ \frac{1}{4}K_{(0,1)}\eta_{\mu\nu}\partial_{\rho}\psi\partial^{\rho}\psi - \frac{1}{2}K_{(0,1)}\partial_{\mu}\psi\partial_{\nu}\psi\,,\\
\mathcal{G}^3_{\mu\nu} &= G_{3(1,0)}\partial_{\mu}\psi\partial_{\nu}\psi - \frac{1}{2}G_{3(1,0)}\eta_{\mu\nu}\partial_{\rho}\psi\partial^{\rho}\psi\,,\\
\mathcal{G}^4_{\mu\nu} &= G_{4(0,0)}G_{\mu\nu}[h^1] + G_{4(1,0)}\eta_{\mu\nu}\square\psi - G_{4(1,0)}\partial_{\mu}\partial_{\nu}\psi + G_{4(1,0)}G_{\mu\nu}[h^1]\psi + G_{4(0,0)}G_{\mu\nu}[h^2]\nonumber\\
&\phantom{=}+ 2G_{4(2,0)}\eta_{\mu\nu}\psi\square\psi + G_{4(1,0)}h_{\mu\nu}\square\psi + G_{4(1,0)}\Gamma^{\rho}{}_{\mu\nu}[h^1]\partial_{\rho}\psi - G_{4(1,0)}\eta_{\mu\nu}\eta^{\rho\sigma}\Gamma^\tau{}_{\rho\sigma}[h^1]\partial_{\tau}\psi\nonumber\\
&\phantom{=}- 2G_{4(2,0)}\psi\partial_{\mu}\partial_{\nu}\psi - G_{4(1,0)}\eta_{\mu\nu}h_{\rho\sigma}\partial^{\rho}\partial^{\sigma}\psi - G_{4(0,1)}\square\psi\partial_{\mu}\partial_{\nu}\psi + G_{4(0,1)}\partial_{\rho}\partial_{\mu}\psi\partial^{\rho}\partial_{\nu}\psi\nonumber\\
&\phantom{=}+ 2G_{4(2,0)}\eta_{\mu\nu}\partial_{\rho}\psi\partial^{\rho}\psi + \frac{1}{2}G_{4(0,1)}\eta_{\mu\nu}\left[(\square\psi)^2 - \partial_{\rho}\partial_{\sigma}\psi\partial^{\rho}\partial^{\sigma}\psi\right] - 2G_{4(2,0)}\partial_{\mu}\psi\partial_{\nu}\psi\,,\\
\mathcal{G}^5_{\mu\nu} &= G_{5(1,0)}\square\psi\partial_{\mu}\partial_{\nu}\psi - G_{5(1,0)}\partial_{\rho}\partial_{\mu}\psi\partial^{\rho}\partial_{\nu}\psi - \frac{1}{2}G_{5(1,0)}\eta_{\mu\nu}\left[(\square\psi)^2 - \partial_{\rho}\partial_{\sigma}\psi\partial^{\rho}\partial^{\sigma}\psi\right]\,,
\end{align}
\end{subequations}
for \(\mathcal{G}^i_{\mu\nu}\),
\begin{subequations}\label{eqn:pexpansion}
\begin{align}
P^2_{\phi} &= K_{(1,0)} + 2K_{(2,0)}\psi + 3K_{(3,0)}\psi^2 - \frac{1}{2}K_{(1,1)}\partial_{\rho}\psi\partial^{\rho}\psi\,,\\
P^3_{\phi} &= 2G_{3(2,0)}\partial_{\rho}\psi\partial^{\rho}\psi\,,\\
P^4_{\phi} &= G_{4(1,0)}R[h^1] + 2G_{4(2,0)}R[h^1]\psi + G_{4(1,0)}R[h^2] + G_{4(1,1)}\left[(\square\psi)^2 - \partial_{\rho}\partial_{\sigma}\psi\partial^{\rho}\partial^{\sigma}\psi\right]\,,\\
P^5_{\phi} &= 0\,,
\end{align}
\end{subequations}
for \(P^i_{\phi}\) and
\begin{subequations}\label{eqn:jexpansion}
\begin{align}
J^2_{\mu} &= -K_{(0,1)}\partial_{\mu}\psi - K_{(1,1)}\psi\partial_{\mu}\psi\,,\\
J^3_{\mu} &= 2G_{3(1,0)}\partial_{\mu}\psi + 4G_{3(2,0)}\psi\partial_{\mu}\psi + \frac{1}{2}G_{3(0,1)}\left[2\square\psi\partial_{\mu}\psi - \partial_{\mu}(\partial_{\rho}\psi\partial^{\rho}\psi)\right]\,,\\
J^4_{\mu} &= 2G_{4(0,1)}G_{\mu\nu}[h^1]\partial^{\nu}\psi - G_{4(1,1)}\left[2\square\psi\partial_{\mu}\psi - \partial_{\mu}(\partial_{\rho}\psi\partial^{\rho}\psi)\right]\,,\\
J^5_{\mu} &= -2G_{5(1,0)}G_{\mu\nu}[h^1]\partial^{\nu}\psi\,,
\end{align}
\end{subequations}
for \(J^i_{\mu}\) up to the quadratic order in \(h_{\mu\nu}\) and \(\psi\). Note that here we have changed our notation from the one we used in the previous section. From this section to the end of section~\ref{sec:mlfullppn}, where we discuss perturbations around a flat background, \(\square = \eta^{\mu\nu}\partial_{\mu}\partial_{\nu}\) denotes the flat Minkowski d'Alembert operator and indices are raised and lowered using the flat metric \(\eta_{\mu\nu}\). Further, we have introduced the notation \(F[h^n]\) for the term which is of order \(n\) in the expansion of \(F\) with respect to the metric perturbation \(h_{\mu\nu}\). For our calculation we further need the trace-reversed terms \(\mathcal{R}^i_{\mu\nu}\) defined in equation~\eqref{eqn:trcurv} and the divergences \(\nabla^{\mu}J^i_{\mu}\). From the expansions~\eqref{eqn:gexpansion} and~\eqref{eqn:jexpansion} one easily derives the expansions
\begin{subequations}\label{eqn:rexpansion}
\begin{align}
\mathcal{R}^2_{\mu\nu} &= \frac{1}{2}K_{(0,0)}\eta_{\mu\nu} + \frac{1}{2}K_{(0,0)}h_{\mu\nu} + \frac{1}{2}K_{(1,0)}\eta_{\mu\nu}\psi + \frac{1}{2}K_{(1,0)}h_{\mu\nu}\psi + \frac{1}{2}K_{(2,0)}\eta_{\mu\nu}\psi^2\nonumber\\
&\phantom{=}- \frac{1}{2}K_{(0,1)}\partial_{\mu}\psi\partial_{\nu}\psi\,,\\
\mathcal{R}^3_{\mu\nu} &= G_{3(1,0)}\partial_{\mu}\psi\partial_{\nu}\psi\,,\\
\mathcal{R}^4_{\mu\nu} &= G_{4(0,0)}R_{\mu\nu}[h^1] - \frac{1}{2}G_{4(1,0)}\eta_{\mu\nu}\square\psi - G_{4(1,0)}\partial_{\mu}\partial_{\nu}\psi + G_{4(1,0)}R_{\mu\nu}[h^1]\psi + G_{4(0,0)}R_{\mu\nu}[h^2]\nonumber\\
&\phantom{=}- G_{4(2,0)}\eta_{\mu\nu}\psi\square\psi - \frac{1}{2}G_{4(1,0)}h_{\mu\nu}\square\psi + \frac{1}{2}G_{4(1,0)}\eta_{\mu\nu}h_{\rho\sigma}\partial^{\rho}\partial^{\sigma}\psi\nonumber\\
&\phantom{=}- 2G_{4(2,0)}\psi\partial_{\mu}\partial_{\nu}\psi + G_{4(1,0)}\Gamma^{\rho}{}_{\mu\nu}[h^1]\partial_{\rho}\psi - G_{4(0,1)}\square\psi\partial_{\mu}\partial_{\nu}\psi + G_{4(0,1)}\partial_{\rho}\partial_{\mu}\psi\partial^{\rho}\partial_{\nu}\psi\nonumber\\
&\phantom{=}- G_{4(2,0)}\eta_{\mu\nu}\partial_{\rho}\psi\partial^{\rho}\psi - 2G_{4(2,0)}\partial_{\mu}\psi\partial_{\nu}\psi + \frac{1}{2}G_{4(1,0)}\eta_{\mu\nu}\eta^{\rho\sigma}\Gamma^\tau{}_{\rho\sigma}[h^1]\partial_{\tau}\psi\,,\\
\mathcal{R}^5_{\mu\nu} &= G_{5(1,0)}\square\psi\partial_{\mu}\partial_{\nu}\psi - G_{5(1,0)}\partial_{\rho}\partial_{\mu}\psi\partial^{\rho}\partial_{\nu}\psi\,,
\end{align}
\end{subequations}
for \(\mathcal{R}^i_{\mu\nu}\) and
\begin{subequations}\label{eqn:djexpansion}
\begin{align}
\nabla^{\mu}J^2_{\mu} &= -K_{(0,1)}\square\psi + K_{(0,1)}h_{\mu\nu}\partial^{\mu}\partial^{\nu}\psi + K_{(0,1)}\eta^{\mu\nu}\Gamma^{\rho}{}_{\mu\nu}\partial_{\rho}\psi - K_{(1,1)}\psi\square\psi - K_{(1,1)}\partial_{\rho}\psi\partial^{\rho}\psi\,,\\
\nabla^{\mu}J^3_{\mu} &= 2G_{3(1,0)}\square\psi - 2G_{3(1,0)}h_{\mu\nu}\partial^{\mu}\partial^{\nu}\psi - 2G_{3(1,0)}\eta^{\mu\nu}\Gamma^{\rho}{}_{\mu\nu}\partial_{\rho}\psi + 4G_{3(2,0)}\psi\square\psi\nonumber\\
&\phantom{=}+ 4G_{3(2,0)}\partial_{\rho}\psi\partial^{\rho}\psi + G_{3(0,1)}\left[(\square\psi)^2 - \partial_{\rho}\partial_{\sigma}\psi\partial^{\rho}\partial^{\sigma}\psi\right]\,,\\
\nabla^{\mu}J^4_{\mu} &= 2G_{4(0,1)}G_{\mu\nu}[h^1]\partial^{\mu}\partial^{\nu}\psi - 2G_{4(1,1)}\left[(\square\psi)^2 - \partial_{\rho}\partial_{\sigma}\psi\partial^{\rho}\partial^{\sigma}\psi\right]\,,\\
\nabla^{\mu}J^5_{\mu} &= -2G_{5(1,0)}G_{\mu\nu}[h^1]\partial^{\mu}\partial^{\nu}\psi\,.
\end{align}
\end{subequations}
for \(\nabla^{\mu}J^i_{\mu}\). From these expressions we see that the only Taylor coefficients relevant for our discussion will be
\begin{gather}
K_{(0,0)}\,, \quad K_{(1,0)}\,, \quad K_{(2,0)}\,, \quad K_{(3,0)}\,, \quad K_{(0,1)}\,, \quad K_{(1,1)}\,, \quad G_{3(1,0)}\,, \quad G_{3(2,0)}\,,\nonumber\\
G_{3(0,1)}\,, \quad G_{4(0,0)}\,, \quad G_{4(1,0)}\,, \quad G_{4(2,0)}\,, \quad G_{4(0,1)}\,, \quad G_{4(1,1)}\,, \quad G_{5(1,0)}\,.
\end{gather}
All other terms in the Taylor expansion would lead to corrections of at least cubic order in the field perturbations.

The perturbative expansions of the field equations around a fixed background spacetime displayed in this section will be a central ingredient for the calculation presented in this article. The second main ingredient will be the parameterized post-Newtonian formalism, which is built upon these perturbative expansions and will be discussed in the following section.

\section{Post-Newtonian approximation}\label{sec:ppn}
The main tool we use in this article is the parameterized post-Newtonian (PPN) formalism~\cite{Will:1993ns,Will:2014xja}, which we briefly review in this section in the context of the given scalar-tensor theory of gravity. The key idea of the PPN formalism is the assumption that the matter which acts as the source of the gravitational field is given by a perfect fluid, whose velocity in a particular, fixed frame of reference is small, measured in units of the speed of light, and that all physical quantities relevant for the solution of the gravitational field equations can be expanded in orders of this velocity. We will now show how this expansion in velocity orders proceeds for the quantities we need in our calculation in the following sections.

The starting point of our calculation is the energy-momentum tensor of a perfect fluid with rest energy density \(\rho\), specific internal energy \(\Pi\), pressure \(p\) and four-velocity \(u^{\mu}\), which takes the form
\begin{equation} \label{eqn:tmunu}
T^{\mu\nu} = (\rho + \rho\Pi + p)u^{\mu}u^{\nu} + pg^{\mu\nu}\,.
\end{equation}
The four-velocity \(u^{\mu}\) is normalized by the metric \(g_{\mu\nu}\), so that \(u^{\mu}u^{\nu}g_{\mu\nu} = -1\). We will now expand all dynamical quantities in orders \(\mathcal{O}(n) \propto |\vec{v}|^n\) of the velocity \(v^{i} = u^{i}/u^0\) of the source matter in a given frame of reference, starting with the field variables. For the metric \(g_{\mu\nu}\) this is an expansion around a flat Minkowski background,
\begin{equation}\label{eqn:metricexp}
g_{\mu\nu} = \eta_{\mu\nu} + h_{\mu\nu} = \eta_{\mu\nu} + h^{(1)}_{\mu\nu} + h^{(2)}_{\mu\nu} + h^{(3)}_{\mu\nu} + h^{(4)}_{\mu\nu} + \mathcal{O}(5)\,,
\end{equation}
while the scalar field \(\phi\) is expanded around its cosmological background value,
\begin{equation}\label{eqn:scalarexp}
\phi = \Phi + \psi = \Phi + \psi^{(1)} + \psi^{(2)} + \psi^{(3)} + \psi^{(4)} + \mathcal{O}(5)\,.
\end{equation}
Here each term \(h^{(n)}_{\mu\nu}\) resp. \(\psi^{(n)}\) is of order \(\mathcal{O}(n)\). In order to describe the motion of test bodies in the lowest post-Newtonian approximation an expansion up to the fourth velocity order \(\mathcal{O}(4)\) is sufficient. A detailed analysis shows that not all components of the metric and the scalar field need to be expanded to the fourth velocity order, while others vanish due to Newtonian energy conservation or time reversal symmetry. The only relevant, non-vanishing components of the field variables are given by
\begin{equation}\label{eqn:ppnfields}
h^{(2)}_{00}\,, \quad h^{(2)}_{ij}\,, \quad h^{(3)}_{0j}\,, \quad h^{(4)}_{00}\,, \quad \psi^{(2)}\,, \quad \psi^{(4)}\,.
\end{equation}
In order to determine these components for a given matter source we must assign velocity orders also to the rest mass density, specific internal energy and pressure of the perfect fluid. Based on their orders of magnitude in the solar system one assigns velocity orders \(\mathcal{O}(2)\) to \(\rho\) and \(\Pi\) and \(\mathcal{O}(4)\) to \(p\). The energy-momentum tensor~\eqref{eqn:tmunu} can then be expanded in the form
\begin{subequations}\label{eqn:energymomentum}
\begin{eqnarray}
T_{00} &=& \rho\left(1 + \Pi + v^2 - h^{(2)}_{00}\right) + \mathcal{O}(6)\,,\\
T_{0j} &=& -\rho v_j + \mathcal{O}(5)\,,\\
T_{ij} &=& \rho v_iv_j + p\delta_{ij} + \mathcal{O}(6)\,.
\end{eqnarray}
\end{subequations}
For later use we also expand the trace-reversed energy momentum tensor introduced in the field equations~\eqref{eqn:trfieldeqns} into terms of velocity orders and obtain the expressions
\begin{subequations}\label{eqn:trenergymomentum}
\begin{align}
\bar{T}_{00} &= \frac{1}{2}\rho + \frac{1}{2}\rho\Pi + \rho v^2 - \frac{1}{2}\rho h^{(2)}_{00} + \frac{3}{2}p + \mathcal{O}(6)\,,\\
\bar{T}_{0j} &= -\rho v_j + \mathcal{O}(5)\,,\\
\bar{T}_{ij} &= \frac{1}{2}\rho\delta_{ij} + \frac{1}{2}\rho\Pi\delta_{ij} + \rho v_iv_j + \frac{1}{2}\rho h^{(2)}_{ij} - \frac{1}{2}p\delta_{ij} + \mathcal{O}(6)\,.
\end{align}
\end{subequations}
We further assume that the gravitational field is quasi-static, so that changes are only induced by the motion of the source matter. Time derivatives \(\partial_0\) of the metric components and other fields are therefore weighted with an additional velocity order \(\mathcal{O}(1)\).

In order to solve the gravitational field equations, which inherit a gauge symmetry from the diffeomorphism invariance of the gravitational action, we finally need to fix a gauge for the metric tensor. A useful choice for the class of scalar-tensor theories we consider can be constructed in analogy to the gauge condition introduced in~\cite{Nutku:1969} and takes the form
\begin{equation}
h_{ij,j} - \frac{1}{2}h_{jj,i} + \frac{1}{2}h_{00,i} = \frac{G_{4(1,0)}}{G_{4(0,0)}}\psi_{,i}\,, \quad h_{0j,j} - \frac{1}{2}h_{jj,0} = \frac{G_{4(1,0)}}{G_{4(0,0)}}\psi_{,0}\,.
\end{equation}
In this gauge the Ricci tensor up to the required order takes the form
\begin{subequations}\label{eqn:riccigauge}
\begin{align}
\label{eqn:r00gauge}
R_{00} &= -\frac{1}{2}h^{(2)}_{00,kk} - \frac{1}{2}h^{(4)}_{00,kk} + \frac{G_{4(1,0)}}{G_{4(0,0)}}\psi^{(2)}_{,00}\\
&\phantom{=}+ \frac{G_{4(1,0)}}{2G_{4(0,0)}}h^{(2)}_{00,j}\psi^{(2)}_{,j} - \frac{1}{2}h^{(2)}_{00,j}h^{(2)}_{00,j} + \frac{1}{2}h^{(2)}_{jk}h^{(2)}_{00,jk} + \mathcal{O}(6)\,,\nonumber\\
R_{0j} &= -\frac{1}{2}h^{(3)}_{0j,kk} - \frac{1}{4}h^{(2)}_{00,0j} + \frac{G_{4(1,0)}}{G_{4(0,0)}}\psi^{(2)}_{,0j} + \mathcal{O}(5)\,,\label{eqn:r0jgauge}\\
R_{ij} &= -\frac{1}{2}h^{(2)}_{ij,kk} + \frac{G_{4(1,0)}}{G_{4(0,0)}}\psi^{(2)}_{,ij} + \mathcal{O}(4)\,.\label{eqn:rijgauge}
\end{align}
\end{subequations}

We now have all necessary tools at hand in order to derive the post-Newtonian limit of Horndeski's gravity theory. Using the expansions~\eqref{eqn:metricexp} and~\eqref{eqn:scalarexp} we can derive the post-Newtonian gravitational field equations from the expansions shown in section~\ref{sec:pertexp}, keeping only the terms~\eqref{eqn:ppnfields}. We further apply the chosen gauge by inserting the gauge fixed Ricci tensor~\eqref{eqn:riccigauge} into the resulting equations. Finally, we can solve these equations following increasing velocity orders. This will be done first under the assumption of a static, spherically symmetric gravitational field generated by a single point mass in the following section, following the approach detailed in~\cite{Hohmann:2013rba}. For a more restricted class of theories, corresponding to a massless scalar field, we can go beyond this assumption and fully solve the post-Newtonian gravitational field equations, and will do so in section~\ref{sec:mlfullppn}.

\section{Static spherically symmetric solution}\label{sec:sphersym}
We will now construct a solution to the post-Newtonian gravitational field equations derived from the perturbative expansions detailed in the previous sections. The central assumption in this section will be that the source of the gravitational field is given by a single point mass, so that the gravitational field is static and spherically symmetric. The general form of this field is explained in section~\ref{subsec:gensssol}. The solution will yield three physical quantities which we will subsequently determine: the effective gravitational constant \(G_{\text{eff}}\) in section~\ref{subsec:newton} and the PPN parameters \(\gamma\) in section~\ref{subsec:gamma} and \(\beta\) in section~\ref{subsec:beta}. The calculation presented here is carried out in full analogy to the calculation displayed in an earlier work for a more restricted class of scalar-tensor theories~\cite{Hohmann:2013rba}.

\subsection{General form of the solution}\label{subsec:gensssol}
The starting point of our calculation is the assumption that the source of the gravitational field is given by a single point-like mass \(M\), whose energy-momentum tensor is of the form~\eqref{eqn:tmunu} with
\begin{equation}\label{eqn:pointmass}
\rho = M\delta(\vec{x})\,, \quad \Pi = 0\,, \quad p = 0\,, \quad v_i = 0\,.
\end{equation}
This simple matter source induces a static and spherically symmetric metric, which can most easily be expressed using isotropic spherical coordinates. In the rest frame of the gravitating mass we use the ansatz
\begin{subequations}\label{eqn:gmunu}
\begin{align}
g_{00} &= -1 + 2G_{\text{eff}}(r)U(r) - 2G_{\text{eff}}^2(r)\beta(r)U^2(r) + \Phi^{(4)}(r) + \mathcal{O}(6)\,,\label{eqn:g00}\\
g_{0j} &= \mathcal{O}(5)\,,\label{eqn:g0j}\\
g_{ij} &= \left[1 + 2G_{\text{eff}}(r)\gamma(r)U(r)\right]\delta_{ij} + \mathcal{O}(4)\,.\label{eqn:gij}
\end{align}
where \(r\) denotes the radial coordinate and the spherically symmetric, static Newtonian potential \(U(r)\) is given by
\end{subequations}
\begin{equation}\label{eqn:sphernewtpot}
U(r) = \frac{M}{r}\,.
\end{equation}
In the potential \(\Phi^{(4)}\) we collect terms of order \(\mathcal{O}(4)\) which are not of the form \(G_{\text{eff}}^2\beta U^2\), such as the gravitational self-energy. Note that we have changed the normalization of the Newtonian potential~\eqref{eqn:sphernewtpot} in comparison to previous work~\cite{Hohmann:2013rba}, where we used an additional factor \(\kappa^2/8\pi\). Here we have omitted this factor in order to be consistent with the standard normalization given in~\cite{Will:1993ns}, which will be relevant in section~\ref{sec:mlfullppn}.

The metric~\eqref{eqn:gmunu} contains three unknown functions which we need to determine. These are the effective gravitational constant \(G_{\text{eff}}(r)\) and the PPN parameters \(\gamma(r)\) and \(\beta(r)\). The latter two can be defined either as the coefficients of the effective gravitational potential \(U_{\text{eff}} = G_{\text{eff}}U\) as shown in the metric~\eqref{eqn:gmunu} or as the coefficients \(\gamma_{\text{eff}} = G_{\text{eff}}\gamma\) and \(\beta_{\text{eff}} = G_{\text{eff}}^2\beta\) of the Newtonian potential terms \(U\) and \(U^2\). The first definition invokes the interpretation that the measured values of \(\gamma\) and \(\beta\) can be related to the effective gravitational potential \(U_{\text{eff}}\), while the second definition suggests to relate the measured values of \(\gamma_{\text{eff}}\) and \(\beta_{\text{eff}}\) to the Newtonian potential \(U\) of a fixed mass \(M\). We choose the first definition in this article since the mass of the Sun, which dominates the solar system physics, is determined from its gravitational effects on the planetary motions.

The Newtonian potential \(U(r)\) we introduced here is of second velocity order, so that the zeroth velocity order solution is given by the flat Minkowski background \(g_{\mu\nu} = \eta_{\mu\nu}\). For this solution to be consistent with the gravitational field equations, we must check that it solves the zeroth order in their perturbative expansion, which corresponds to the equations of motion for the cosmological background. The corresponding equations follow from our assumption that this background is stationary and are given by
\begin{equation}\label{eqn:bgcondition}
\frac{1}{2}K_{(0,0)}\eta_{\mu\nu} = 0\,, \quad K_{(1,0)} = 0\,.
\end{equation}
As one can see, these are solved only if the Taylor series coefficients \(K_{(0,0)}\) and \(K_{(1,0)}\) vanish. We will therefore restrict ourselves to theories which satisfy these conditions.

\subsection{Newtonian approximation}\label{subsec:newton}
We will now determine the effective gravitational constant \(G_{\text{eff}}(r)\) appearing in the metric solution~\eqref{eqn:gmunu}. The starting point of this calculation is given by the gravitational field equations~\eqref{eqn:fieldeqns} and~\eqref{eqn:trfieldeqns}. At the second velocity order they are given by
\begin{subequations}
\begin{align}
\frac{1}{2}\bar{T}^{(2)}_{00} &= G_{4(0,0)}R^{(2)}_{00} + \frac{1}{2}G_{4(1,0)}\psi^{(2)}_{,ii}\,,\label{eqn:r200}\\
\frac{1}{2}\bar{T}^{(2)}_{ij} &= G_{4(0,0)}R^{(2)}_{ij} - \frac{1}{2}G_{4(1,0)}\psi^{(2)}_{,kk}\delta_{ij} - G_{4(1,0)}\psi^{(2)}_{,ij}\,,\label{eqn:r2ij}\\
0 &= \left(K_{(0,1)} - 2G_{3(1,0)}\right)\psi^{(2)}_{,ii} + 2K_{(2,0)}\psi^{(2)} + G_{4(1,0)}R^{(2)}\,,\label{eqn:scalar2}
\end{align}
\end{subequations}
where \(R^{(2)}_{\mu\nu}\) is the second velocity order part of the Ricci tensor. Here we already used the restriction \(K_{(0,0)} = K_{(1,0)} = 0\) derived from the zeroth order field equations. The components of the trace-reversed energy-momentum tensor are given by
\begin{equation}
\bar{T}^{(2)}_{00} = \frac{1}{2}\rho\,, \quad \bar{T}^{(2)}_{ij} = \frac{1}{2}\rho\delta_{ij}\,.
\end{equation}
In order to eliminate the second order Ricci scalar \(R^{(2)} = -R^{(2)}_{00} + R^{(2)}_{ii}\) from the third equation~\eqref{eqn:scalar2}, we can take the trace
\begin{equation}\label{eqn:trace2}
\frac{1}{2}\rho = G_{4(0,0)}\left(-R^{(2)}_{00} + R^{(2)}_{ii}\right) - 3G_{4(1,0)}\psi^{(2)}_{,ii}\,,
\end{equation}
over the first two equations~\eqref{eqn:r200} and~\eqref{eqn:r2ij}. This yields the scalar field equation
\begin{equation}
\left(K_{(0,1)} - 2G_{3(1,0)} + 3\frac{G_{4(1,0)}^2}{G_{4(0,0)}}\right)\psi^{(2)}_{,ii} + 2K_{(2,0)}\psi^{(2)} + \frac{G_{4(1,0)}}{2G_{4(0,0)}}\rho = 0\,,
\end{equation}
which has the form of a screened Poisson equation,
\begin{equation}\label{eqn:psi2}
\psi^{(2)}_{,ii} - m_{\psi}^2\psi^{(2)} = -c_{\psi}\rho\,,
\end{equation}
where we introduced the constants
\begin{equation}\label{eqn:mpsi}
m_{\psi} = \sqrt{\frac{-2K_{(2,0)}}{K_{(0,1)} - 2G_{3(1,0)} + 3\frac{G_{4(1,0)}^2}{G_{4(0,0)}}}}\,, \quad c_{\psi} = \frac{G_{4(1,0)}}{2G_{4(0,0)}}\left(K_{(0,1)} - 2G_{3(1,0)} + 3\frac{G_{4(1,0)}^2}{G_{4(0,0)}}\right)^{-1}\,.
\end{equation}
In order for a consistent solution to exist which is compatible with the perturbation ansatz we restrict ourselves to the case \(K_{(2,0)} \leq 0\) and \(K_{(0,1)} - 2G_{3(1,0)} + 3G_{4(1,0)}^2/G_{4(0,0)} > 0\). The solution is then given by
\begin{equation}
\psi^{(2)}(r) = \frac{M}{4\pi r}c_{\psi}e^{-m_{\psi}r}
\end{equation}
for the point mass source~\eqref{eqn:pointmass}.

In the next step we use equation~\eqref{eqn:r200} and insert the second order Ricci tensor \(R^{(2)}_{00}\) from equation~\eqref{eqn:r00gauge}. In the resulting equation for \(h^{(2)}_{00}\) we eliminate the term \(\psi^{(2)}_{,ii}\) using equation~\eqref{eqn:psi2} and finally obtain
\begin{equation}\label{eqn:h200}
h^{(2)}_{00,ii} = c_1\psi^{(2)} - c_2\rho\,,
\end{equation}
where the constants \(c_{1,2}\) are given by
\begin{subequations}
\begin{align}
c_1 &= -2\frac{G_{4(1,0)}K_{(2,0)}}{G_{4(0,0)}}\left(K_{(0,1)} - 2G_{3(1,0)} + 3\frac{G_{4(1,0)}^2}{G_{4(0,0)}}\right)^{-1}\,,\label{eqn:c1def}\\
c_2 &= \frac{1}{G_{4(0,0)}}\left[\frac{1}{2} + \frac{G_{4(1,0)}^2}{2G_{4(0,0)}}\left(K_{(0,1)} - 2G_{3(1,0)} + 3\frac{G_{4(1,0)}^2}{G_{4(0,0)}}\right)^{-1}\right]\,.\label{eqn:c2def}
\end{align}
\end{subequations}
For the point mass source~\eqref{eqn:pointmass} this equation has the solution
\begin{equation}
h^{(2)}_{00}(r) = \frac{M}{4\pi r}\left[c_2 + \frac{c_1c_{\psi}}{m_{\psi}^2}(e^{-m_{\psi}r} - 1)\right]\,.
\end{equation}
By comparison with the metric component~\eqref{eqn:g00} we read off the effective gravitational constant
\begin{equation}\label{eqn:geff}
G_{\text{eff}}(r) = \frac{1}{8\pi}\left[c_2 + \frac{c_1c_{\psi}}{m_{\psi}^2}(e^{-m_{\psi}r} - 1)\right]\,.
\end{equation}
In order to interpret this result for \(G_{\text{eff}}\) as an effective gravitational constant we need to choose an experiment in which the gravitational interaction takes place at a constant scale \(r = r_0\). We can then choose units in which \(G_{\text{eff}}(r_0) = 1\). This corresponds to a rescaling of the parameter functions \(K, G_3, G_4, G_5\). However, we cannot make this choice globally, and hence cannot remove the factor \(G_{\text{eff}}(r)\) from the metric~\eqref{eqn:gmunu} by a choice of units in which \(G_{\text{eff}} \equiv 1\), as it is conventionally done in the basic PPN formalism~\cite{Will:1993ns}. This is the reason for the ambiguity in the definition of the PPN parameters \(\gamma\) and \(\beta\) we discussed above.

\subsection{PPN parameter $\gamma(r)$}\label{subsec:gamma}
We now come to the calculation of the PPN parameter \(\gamma\), which can be read off from the spatial perturbation component \(h^{(2)}_{ij}\), as it appears in the corresponding term in the metric~\eqref{eqn:gij}. For this purpose we use the field equation~\eqref{eqn:r2ij} and insert the second order Ricci tensor \(R^{(2)}_{ij}\) from equation~\eqref{eqn:rijgauge}. As we did in the previous section when calculating the effective gravitational constant, we eliminate the term \(\psi^{(2)}_{,ii}\) using equation~\eqref{eqn:psi2} and finally obtain
\begin{equation}\label{eqn:h2ij}
h^{(2)}_{ij,kk} = \left(c_3\psi^{(2)} - c_4\rho\right)\delta_{ij}\,,
\end{equation}
where the constants \(c_{3,4}\) are given by
\begin{subequations}
\begin{align}
c_3 &= 2\frac{G_{4(1,0)}K_{(2,0)}}{G_{4(0,0)}}\left(K_{(0,1)} - 2G_{3(1,0)} + 3\frac{G_{4(1,0)}^2}{G_{4(0,0)}}\right)^{-1}\,,\label{eqn:c3def}\\
c_4 &= \frac{1}{G_{4(0,0)}}\left[\frac{1}{2} - \frac{G_{4(1,0)}^2}{2G_{4(0,0)}}\left(K_{(0,1)} - 2G_{3(1,0)} + 3\frac{G_{4(1,0)}^2}{G_{4(0,0)}}\right)^{-1}\right]\,.\label{eqn:c4def}
\end{align}
\end{subequations}
This equation has the solution
\begin{equation}
h^{(2)}_{ij}(r) = \frac{M}{4\pi r}\left[c_4 + \frac{c_3c_{\psi}}{m_{\psi}^2}(e^{-m_{\psi}r} - 1)\right]\delta_{ij}
\end{equation}
for the point mass source~\eqref{eqn:pointmass}. By comparison with the metric component~\eqref{eqn:gij} one reads off the PPN parameter
\begin{equation}\label{eqn:gamma}
\gamma(r) = \frac{c_4 + \frac{c_3c_{\psi}}{m_{\psi}^2}(e^{-m_{\psi}r} - 1)}{c_2 + \frac{c_1c_{\psi}}{m_{\psi}^2}(e^{-m_{\psi}r} - 1)} = \frac{2\omega + 3 - e^{-m_{\psi}r}}{2\omega + 3 + e^{-m_{\psi}r}}\,,
\end{equation}
where the constant \(\omega\) is given by
\begin{equation}\label{eqn:omega}
\omega = \frac{G_{4(0,0)}}{2G_{4(1,0)}^2}\left(K_{(0,1)} - 2G_{3(1,0)}\right)\,.
\end{equation}
The result reproduces a previously derived result for the PPN parameter \(\gamma(r)\) for a more restricted class of scalar-tensor theories of gravity~\cite{Hohmann:2013rba}. It thus also yields analogous limiting cases, which are obtained as follows. In the limit \(m_{\psi} \to 0\) and fixed finite \(\omega\), the PPN parameter \(\gamma\) approaches the known value
\begin{equation}
\gamma = \frac{\omega + 1}{\omega + 2}
\end{equation}
for scalar-tensor gravity with a massless scalar field~\cite{Nordtvedt:1970uv}. In the limit \(\omega \to \infty\) we find the limiting value \(\gamma = 1\), independent of \(m_{\psi}\). The same value \(\gamma = 1\) is also approached in the limiting case of a massive scalar field with \(m_{\psi}r \gg 1\).

\subsection{PPN parameter $\beta(r)$}\label{subsec:beta}
We finally come to the calculation of the PPN parameter \(\beta\), which is read off from the component \(h^{(4)}_{00}\), which follows from the metric term~\eqref{eqn:g00}. Since the field equations at the third velocity order
\begin{equation}
\frac{1}{2}\bar{T}^{(3)}_{0i} = G_{4(0,0)}R^{(3)}_{0i} - G_{4(1,0)}\psi^{(2)}_{,0i}\label{eqn:r30i}
\end{equation}
are solved identically for the static, spherically symmetric solution we consider here, we can directly proceed with the fourth order field equations. These take the form
\begin{subequations}
\begin{align}
\frac{1}{2}\bar{T}^{(4)}_{00} &= G_{4(0,0)}R^{(4)}_{00} + \frac{1}{2}G_{4(1,0)}\psi^{(4)}_{,ii} - \frac{1}{2}K_{(2,0)}\left(\psi^{(2)}\right)^2 - \frac{3}{2}G_{4(1,0)}\psi^{(2)}_{,00} + G_{4(1,0)}R^{(2)}_{00}\psi^{(2)}\nonumber\\
&\phantom{=}+ G_{4(2,0)}\psi^{(2)}_{,ii}\psi^{(2)} - \frac{1}{2}G_{4(1,0)}h^{(2)}_{00}\psi^{(2)}_{,ii} - \frac{1}{2}G_{4(1,0)}h^{(2)}_{ij}\psi^{(2)}_{,ij} - \frac{1}{2}G_{4(1,0)}h^{(2)}_{00,i}\psi^{(2)}_{,i}\nonumber\\
&\phantom{=}+ G_{4(2,0)}\psi^{(2)}_{,i}\psi^{(2)}_{,i} - \frac{1}{2}G_{4(1,0)}\left(h^{(2)}_{ij,i} - \frac{1}{2}h^{(2)}_{ii,j} + \frac{1}{2}h^{(2)}_{00,j}\right)\psi^{(2)}_{,j}\,,\label{eqn:r400}\\
\frac{1}{2}\bar{T}^{(4)}_{ij} &= G_{4(0,0)}R^{(4)}_{ij} - \frac{1}{2}G_{4(1,0)}\psi^{(4)}_{,kk}\delta_{ij} - G_{4(1,0)}\psi^{(4)}_{,ij} + \frac{1}{2}K_{(2,0)}\left(\psi^{(2)}\right)^2\delta_{ij}\nonumber\\
&\phantom{=}+ \left(G_{3(1,0)} - \frac{1}{2}K_{(0,1)} - 2G_{4(2,0)}\right)\psi^{(2)}_{,i}\psi^{(2)}_{,j} + \frac{1}{2}G_{4(1,0)}\psi^{(2)}_{,00}\delta_{ij} + G_{4(1,0)}R^{(2)}_{ij}\psi^{(2)}\nonumber\\
&\phantom{=}- G_{4(2,0)}\psi^{(2)}_{,kk}\psi^{(2)}\delta_{ij} - \frac{1}{2}G_{4(1,0)}h^{(2)}_{ij}\psi^{(2)}_{,kk} + \frac{1}{2}G_{4(1,0)}h^{(2)}_{kl}\psi^{(2)}_{,kl}\delta_{ij} - 2G_{4(2,0)}\psi^{(2)}_{,ij}\psi^{(2)}\nonumber\\
&\phantom{=}+ \frac{1}{2}G_{4(1,0)}\left(h^{(2)}_{ik,j} + h^{(2)}_{jk,i} - h^{(2)}_{ij,k}\right)\psi^{(2)}_{,k} + \frac{1}{2}G_{4(1,0)}\left(h^{(2)}_{kl,k} - \frac{1}{2}h^{(2)}_{kk,l} + \frac{1}{2}h^{(2)}_{00,l}\right)\psi^{(2)}_{,l}\delta_{ij}\nonumber\\
&\phantom{=}- G_{4(2,0)}\psi^{(2)}_{,k}\psi^{(2)}_{,k}\delta_{ij} + \left(G_{5(1,0)} - G_{4(0,1)}\right)\left(\psi^{(2)}_{,ij}\psi^{(2)}_{,kk} - \psi^{(2)}_{,ik}\psi^{(2)}_{,jk}\right)\,,\label{eqn:r4ij}\\
0 &= \left(K_{(0,1)} - 2G_{3(1,0)}\right)\psi^{(4)}_{,ii} + 2K_{(2,0)}\psi^{(4)} + G_{4(1,0)}R^{(4)} + 3K_{(3,0)}\left(\psi^{(2)}\right)^2\nonumber\\
&\phantom{=}+ \left(2G_{3(1,0)} - K_{(0,1)}\right)\left[\psi^{(2)}_{,00} + h^{(2)}_{ij}\psi^{(2)}_{,ij} + \left(h^{(2)}_{ij,i} - \frac{1}{2}h^{(2)}_{ii,j} + \frac{1}{2}h^{(2)}_{00,j}\right)\psi^{(2)}_{,j}\right]\nonumber\\
&\phantom{=}+ 2G_{4(2,0)}R^{(2)}\psi^{(2)} + \left(\frac{1}{2}K_{(1,1)} - 2G_{3(2,0)}\right)\left(2\psi^{(2)}_{,ii}\psi^{(2)} + \psi^{(2)}_{,i}\psi^{(2)}_{,i}\right)\nonumber\\
&\phantom{=}+ \left(3G_{4(1,1)} - G_{3(0,1)}\right)\left[\left(\psi^{(2)}_{,ii}\right)^2 - \psi^{(2)}_{,ij}\psi^{(2)}_{,ij}\right] + 2\left(G_{5(1,0)} - G_{4(0,1)}\right)G^{(2)}_{ij}\psi^{(2)}_{,ij}\,,\label{eqn:scalar4}
\end{align}
\end{subequations}
where the components of the trace-reversed energy-momentum tensor read
\begin{equation}
\bar{T}^{(4)}_{00} = \frac{1}{2}\rho\Pi + \rho v^2 - \frac{1}{2}\rho h^{(2)}_{00} + \frac{3}{2}p\,, \quad \bar{T}^{(4)}_{ij} = \frac{1}{2}\rho\Pi\delta_{ij} + \rho v_iv_j + \frac{1}{2}\rho h^{(2)}_{ij} - \frac{1}{2}p\delta_{ij}\,.
\end{equation}
We can eliminate the fourth order Ricci scalar \(R^{(4)} = -R^{(4)}_{00} + R^{(4)}_{ii} - h^{(2)}_{00}R^{(2)}_{00} - h^{(2)}_{ij}R^{(2)}_{ij}\) from the fourth order scalar equation~\eqref{eqn:scalar4} using a suitable linear combination of the fourth order equations~\eqref{eqn:r400} and~\eqref{eqn:r4ij} and the second order equations~\eqref{eqn:r200} and~\eqref{eqn:r2ij}, which reads
\begin{equation}\label{eqn:trace4}
\begin{split}
\frac{1}{2}\rho\Pi - \frac{3}{2}p &= G_{4(0,0)}\left(R^{(4)}_{ii} - R^{(4)}_{00} - h^{(2)}_{ij}R^{(2)}_{ij} - h^{(2)}_{00}R^{(2)}_{00}\right) - 3G_{4(1,0)}\psi^{(4)}_{,ii} + 3G_{4(1,0)}\psi^{(2)}_{,00}\\
&\phantom{=}+ 3G_{4(1,0)}\left(h^{(2)}_{ij,i} - \frac{1}{2}h^{(2)}_{ii,j} + \frac{1}{2}h^{(2)}_{00,j}\right)\psi^{(2)}_{,j} + 3G_{4(1,0)}h^{(2)}_{ij}\psi^{(2)}_{,ij} + 2K_{(2,0)}\left(\psi^{(2)}\right)^2\\
&\phantom{=}+ \left(G_{3(1,0)} - \frac{1}{2}K_{(0,1)} - 6G_{4(2,0)}\right)\psi^{(2)}_{,i}\psi^{(2)}_{,i} - 6G_{4(2,0)}\psi^{(2)}_{,ii}\psi^{(2)}\\
&\phantom{=}+ G_{4(1,0)}\left(R^{(2)}_{ii} - R^{(2)}_{00}\right)\psi^{(2)}+ \left(G_{5(1,0)} - G_{4(0,1)}\right)\left[\left(\psi^{(2)}_{,ii}\right)^2 - \psi^{(2)}_{,ij}\psi^{(2)}_{,ij}\right]\,.
\end{split}
\end{equation}
The resulting equation for the scalar field \(\psi^{(4)}\) finally takes the form
\begin{equation}\label{eqn:psi4}
\begin{split}
\psi^{(4)}_{,ii} - m_{\psi}^2\psi^{(4)} &= d_1\psi^{(2)}_{,00} + d_2\left(\psi^{(2)}\right)^2 + d_3\psi^{(2)}_{,i}\psi^{(2)}_{,i} + d_4\psi^{(2)}_{,ii}\psi^{(2)} + d_5\left(\psi^{(2)}_{,ii}\right)^2\\
&\phantom{=}+ d_6\psi^{(2)}_{,ij}\psi^{(2)}_{,ij} + d_7\left(h^{(2)}_{ij,i} - \frac{1}{2}h^{(2)}_{ii,j} + \frac{1}{2}h^{(2)}_{00,j}\right)\psi^{(2)}_{,j} + d_8R^{(2)}\psi^{(2)}\\
&\phantom{=}+ d_9h^{(2)}_{ij}\psi^{(2)}_{,ij} + d_{10}G^{(2)}_{ij}\psi^{(2)}_{,ij} + d_{11}\rho\Pi + d_{12}p\,,
\end{split}
\end{equation}
where the constants \(d_1, \ldots, d_{12}\) are listed in equation~\eqref{eqn:psi4coeff} in appendix~\ref{sec:expcoeff}. Terms of the forms \(\left(\psi^{(2)}_{,ii}\right)^2\), \(\psi^{(2)}_{,ij}\psi^{(2)}_{,ij}\) and \(G^{(2)}_{ij}\psi^{(2)}_{,ij}\) contain squared second derivatives of the Newtonian potential, and thus squares of the matter density. These terms do not appear in the standard PPN formalism and their influence on the current methods to measure \(\beta\) and other PPN parameters must be determined by a separate phenomenological discussion. We will not enter this discussion here, and therefore restrict ourselves to gravity theories in which the free functions in the action~\eqref{eqn:action} are chosen so that \(d_5 = d_6 = d_{10} = 0\). By comparison with their values listed in equation~\eqref{eqn:psi4coeff} this corresponds to the restrictions
\begin{equation}\label{eqn:squarerestrict}
G_{3(0,1)} = 3G_{4(1,1)}\,, \quad G_{4(0,1)} = G_{5(1,0)}
\end{equation}
on the Taylor expansion coefficients of the functions \(G_3, G_4, G_5\). We now insert the point mass source~\eqref{eqn:pointmass} and the already determined solution for \(\psi^{(2)}\), \(h^{(2)}_{00}\) and \(h^{(2)}_{ij}\). We further neglect terms of the form \(\rho U\), which correspond to gravitational self-energies and thus contribute only to the term \(\Phi^{(4)}(r)\) in the metric component~\eqref{eqn:g00}. The resulting equation then reads
\begin{equation}\label{eqn:psi42}
\psi^{(4)}_{,ii} - m_{\psi}^2\psi^{(4)} = e_1\frac{e^{-m_{\psi}r}}{r^2} + e_2\frac{e^{-2m_{\psi}r}}{r^2} + e_3\frac{e^{-m_{\psi}r}}{r^3} + e_4\frac{e^{-2m_{\psi}r}}{r^3} + e_5\frac{e^{-m_{\psi}r}}{r^4} + e_6\frac{e^{-2m_{\psi}r}}{r^4}\,,
\end{equation}
where the constants \(e_1, \ldots, e_6\) are listed in equation~\eqref{eqn:psi42coeff} in appendix~\ref{sec:expcoeff}. From this equation we obtain the solution
\begin{equation}\label{eqn:psi4sol}
\begin{split}
\psi^{(4)} &= f_1\frac{e^{-m_{\psi}r}}{r^2} + f_2\frac{e^{-2m_{\psi}r}}{r^2} + f_3\frac{e^{-m_{\psi}r}}{r}\ln(m_{\psi}r) + f_4\frac{e^{-m_{\psi}r}}{r}\mathrm{Ei}(-m_{\psi}r)\\
&\phantom{=}+ f_5\frac{e^{m_{\psi}r}}{r}\mathrm{Ei}(-2m_{\psi}r) + f_6\frac{e^{m_{\psi}r}}{r}\mathrm{Ei}(-3m_{\psi}r)\,,
\end{split}
\end{equation}
with constants \(f_1, \ldots, f_6\) listed in equation~\eqref{eqn:psi4solcoeff} in appendix~\ref{sec:expcoeff}. Here \(\mathrm{Ei}\) denotes the exponential integral, which is defined by
\begin{equation}
\mathrm{Ei}(-x) = -\int_{x}^{\infty} \frac{e^{-t}}{t} dt\,.
\end{equation}
In order to determine \(h^{(4)}_{00}\) we now eliminate the term \(\psi^{(4)}_{,ii}\) from equation~\eqref{eqn:r400} by making use of equation~\eqref{eqn:psi4}. The resulting equation then takes the form
\begin{equation}\label{eqn:h400}
\begin{split}
h^{(4)}_{00,ii} &= q_1\psi^{(2)}_{,00} + q_2\left(\psi^{(2)}\right)^2 + q_3\psi^{(2)}_{,i}\psi^{(2)}_{,i} + q_4\psi^{(2)}_{,ii}\psi^{(2)} + q_5\left(\psi^{(2)}_{,ii}\right)^2 + q_6\psi^{(2)}_{,ij}\psi^{(2)}_{,ij}\\
&\phantom{=}+ q_7\left(h^{(2)}_{ij,i} - \frac{1}{2}h^{(2)}_{ii,j} + \frac{1}{2}h^{(2)}_{00,j}\right)\psi^{(2)}_{,j} + q_8R^{(2)}\psi^{(2)} + q_9h^{(2)}_{ij}\psi^{(2)}_{,ij} + q_{10}G^{(2)}_{ij}\psi^{(2)}_{,ij}\\
&\phantom{=}+ q_{11}h^{(2)}_{00,i}\psi^{(2)}_{,i} + q_{12}h^{(2)}_{00}\psi^{(2)}_{,ii} + q_{13}R^{(2)}_{00}\psi^{(2)} + q_{14}h^{(2)}_{00,i}h^{(2)}_{00,i} + q_{15}h^{(2)}_{ij}h^{(2)}_{00,ij}\\
&\phantom{=}+ q_{16}\psi^{(4)} + q_{17}\rho\Pi + q_{18}\rho v^2 + q_{19}\rho h^{(2)}_{00} + q_{20}p\,,
\end{split}
\end{equation}
where the constants \(q_1, \ldots, q_{20}\) are listed in equation~\eqref{eqn:h400coeff} in appendix~\ref{sec:expcoeff}. As it was also the case in equation~\eqref{eqn:psi4}, we find terms of the forms \(\left(\psi^{(2)}_{,ii}\right)^2\), \(\psi^{(2)}_{,ij}\psi^{(2)}_{,ij}\) and \(G^{(2)}_{ij}\psi^{(2)}_{,ij}\), which do not appear in the standard PPN formalism and which we therefore eliminate by the restriction \(q_5 = q_6 = q_{10} = 0\). A calculation of these coefficients shows that they already vanish as a consequence of the restriction~\eqref{eqn:squarerestrict} we imposed earlier, so that all terms involving squares of second derivatives drop out. Into the remaining equation we insert the point mass~\eqref{eqn:pointmass} and the previously determined solutions for the scalar field and the metric perturbations. Again we neglect all gravitational self-energy terms of the form \(\rho U\). This yields us the equation
\begin{equation}\label{eqn:h4002}
\begin{split}
h^{(4)}_{00,ii} &= s_1\frac{e^{-m_{\psi}r}}{r^2} + s_2\frac{e^{-2m_{\psi}r}}{r^2} + s_3\frac{e^{-m_{\psi}r}}{r^3} + s_4\frac{e^{-2m_{\psi}r}}{r^3} + s_5\frac{e^{-m_{\psi}r}}{r^4} + s_6\frac{e^{-2m_{\psi}r}}{r^4}\\
&\phantom{=}+ s_7\frac{1}{r^4} + s_8\frac{e^{-m_{\psi}r}}{r}\ln(m_{\psi}r) + s_9\frac{e^{-m_{\psi}r}}{r}\mathrm{Ei}(-m_{\psi}r) + s_{10}\frac{e^{m_{\psi}r}}{r}\mathrm{Ei}(-2m_{\psi}r)\\
&\phantom{=}+ s_{11}\frac{e^{m_{\psi}r}}{r}\mathrm{Ei}(-3m_{\psi}r)\,,
\end{split}
\end{equation}
where we used the constants \(s_1, \ldots, s_{11}\) listed in equation~\eqref{eqn:h4002coeff} in appendix~\ref{sec:expcoeff}. The solution is then given by
\begin{equation}\label{eqn:h400sol}
\begin{split}
h^{(4)}_{00} &= u_1\frac{1}{r^2} + u_2\frac{e^{-m_{\psi}r}}{r^2} + u_3\frac{e^{-2m_{\psi}r}}{r^2} + u_4\frac{e^{-m_{\psi}r}}{r} + u_5\frac{e^{-2m_{\psi}r}}{r} + u_6\frac{e^{-m_{\psi}r}}{r}\ln(m_{\psi}r)\\
&\phantom{=}+ u_7\frac{1}{r}\mathrm{Ei}(-m_{\psi}r) + u_8\mathrm{Ei}(-m_{\psi}r) + u_9\frac{e^{-m_{\psi}r}}{r}\mathrm{Ei}(-m_{\psi}r) + u_{10}\frac{1}{r}\mathrm{Ei}(-2m_{\psi}r)\\
&\phantom{=}+ u_{11}\mathrm{Ei}(-2m_{\psi}r) + u_{12}\frac{e^{m_{\psi}r}}{r}\mathrm{Ei}(-2m_{\psi}r) + u_{13}\frac{e^{m_{\psi}r}}{r}\mathrm{Ei}(-3m_{\psi}r)\,,
\end{split}
\end{equation}
with constants \(u_1, \ldots, u_{13}\) listed in equation~\eqref{eqn:h400solcoeff} in appendix~\ref{sec:expcoeff}. By comparison with equation~\eqref{eqn:g00} and after inserting all coefficients listed in appendix~\ref{sec:expcoeff} we finally read off
\begin{equation}\label{eqn:beta}
\begin{split}
\beta(r) &= 1 + \frac{1}{(2\omega + 3 + e^{-m_{\psi}r})^2}\Bigg\{\frac{\omega + \tau - 4\omega\sigma}{2\omega + 3}e^{-2m_{\psi}r}\\
&\phantom{=}+ (2\omega + 3)m_{\psi}r\left[e^{-m_{\psi}r}\ln(m_{\psi}r) - \left(m_{\psi}r + e^{m_{\psi}r}\right)\mathrm{Ei}(-2m_{\psi}r) - \frac{1}{2}e^{-2m_{\psi}r}\right]\\
&\phantom{=}+ \frac{6\mu r + 3(3\omega + \tau + 6\sigma + 3)m_{\psi}^2r}{2(2\omega + 3)m_{\psi}}\left[e^{m_{\psi}r}\mathrm{Ei}(-3m_{\psi}r) - e^{-m_{\psi}r}\mathrm{Ei}(-m_{\psi}r)\right]\Bigg\}\,,
\end{split}
\end{equation}
where \(\omega\) is given by equation~\eqref{eqn:omega} and we further introduced the abbreviations
\begin{equation}
\sigma = \frac{G_{4(0,0)}G_{4(2,0)}}{G_{4(1,0)}^2}\,, \quad \tau = \frac{G_{4(0,0)}^2}{2G_{4(1,0)}^3}(K_{(1,1)} - 4G_{3(2,0)})\,, \quad \mu = \frac{G_{4(0,0)}^2K_{(3,0)}}{G_{4(1,0)}^3}\,.\label{eqn:abbreviations}
\end{equation}
We thus see that the result for \(\beta(r)\) has essentially the same structure as a previously found result for a more restricted class of scalar-tensor theories of gravity~\cite{Hohmann:2013rba}. It follows from the asymptotic behavior of the exponential integral in the case \(x \gg 1\),
\begin{equation}
\mathrm{Ei}(-x) \approx \frac{e^{-x}}{x}\left(1 - \frac{1!}{x} + \frac{2!}{x^2} - \frac{3!}{x^3} + \ldots\right)\,,
\end{equation}
that all terms involving \(\sigma\), \(\tau\) or \(\mu\) fall off proportional to \(e^{-2m_{\psi}r}\), and are thus subleading to the terms involving only \(\omega\) and \(m_{\psi}\) which fall off proportional to \(e^{-m_{\psi}r}\). We therefore conclude that at large distances \(m_{\psi}r \gg 1\) from the source the contributions of \(\sigma\), \(\tau\) and \(\mu\) may be neglected. This means in particular that the comparison of \(\gamma(r)\) and \(\beta(r)\) with experiments in the large distance limit detailed in~\cite{Hohmann:2013rba} is valid also in the more general case of Horndeski's gravity theory considered here. We will explain this limit in more detail in section~\ref{sec:obs}.

Again we consider the three limiting cases which we already discussed for \(\gamma\). In the limit \(m_{\psi} \to 0\) and fixed finite \(\omega\) we obtain
\begin{equation}
\beta = 1 + \frac{\omega + \tau - 4\omega\sigma}{(2\omega + 3)(2\omega + 4)^2}\,,
\end{equation}
which essentially reproduces the known result for a massless scalar field~\cite{Nordtvedt:1970uv}. The second case \(\omega \to \infty\) and arbitrary \(m_{\psi}\) yields the limit \(\beta = 1\). We also find the limiting value \(\beta = 1\) in the case \(m_{\psi}r \gg 1\) of a massive scalar field.

This result completes our solution to the post-Newtonian field equations for a static point mass source of gravity. We have calculated the metric up to the first post-Newtonian order as displayed in equation~\eqref{eqn:gmunu}. From our calculation we obtained expressions for the effective gravitational constant~\eqref{eqn:geff} and the PPN parameters \(\gamma\)~\eqref{eqn:gamma} and \(\beta\)~\eqref{eqn:beta}. In the next section we will consider a more restricted class of theories, for which we can solve the post-Newtonian field equations for arbitrary mass distributions and obtain a full set of PPN parameters.

\section{Full set of PPN parameters for a massless scalar field}\label{sec:mlfullppn}
In the previous section we have determined the PPN parameters \(\gamma\) and \(\beta\) from the static, spherically symmetric metric ansatz~\eqref{eqn:gmunu}. For the consistency of this ansatz we had to impose the conditions~\eqref{eqn:bgcondition} and~\eqref{eqn:squarerestrict} on the functions \(K, G_3, G_4, G_5\) in the gravitational action. We have seen that the parameters \(\gamma\) and \(\beta\) depend on the distance between the mass source and the probing test mass, due to the fact that the scalar field acquires a non-vanishing mass~\eqref{eqn:mpsi}. We will now consider a further restricted class of theories in which this mass term vanishes, which is the case if the Taylor expansion coefficient \(K_{(2,0)}\) vanishes. Further, we require that the mass-like (derivative free) term \(\left(\psi^{(2)}\right)^2\) in equations~\eqref{eqn:psi4} and~\eqref{eqn:h400} vanishes, which is achieved by \(K_{(3,0)} = 0\). When this restriction is imposed, it will turn out that the gravitational field equations can be solved for arbitrary matter sources given by the energy-momentum tensor~\eqref{eqn:tmunu}, and that their solution assumes the standard PPN form, from which the full set of ten PPN parameters can be read off~\cite{Will:1993ns}. In order to determine this solution, we will solve the gravitational field equations by increasing velocity orders - the second velocity order in section~\ref{subsec:ppn2}, the third velocity order in section~\ref{subsec:ppn3} and the fourth velocity order in section~\ref{subsec:ppn4}.

\subsection{Second velocity order}\label{subsec:ppn2}
We start by solving the gravitational field equations at the second velocity order following the same steps as in the preceding section, i.e., we first determine \(\psi^{(2)}\), then \(h^{(2)}_{00}\) and finally \(h^{(2)}_{ij}\). For the scalar field we see that equation~\eqref{eqn:psi2} reduces to
\begin{equation}
\psi^{(2)}_{,ii} = -c_{\psi}\rho\,,
\end{equation}
where \(c_{\psi}\) is given by equation~\eqref{eqn:mpsi}, and thus takes the form of an ordinary Poisson equation. The solution is given by
\begin{equation}
\psi^{(2)} = \frac{c_{\psi}}{4\pi}U\,,
\end{equation}
where we have introduced the Newtonian potential
\begin{equation}\label{eqn:newtpot}
U(t,\vec{x}) = \int d^3x'\frac{\rho(t,\vec{x}')}{|\vec{x} - \vec{x}'|}\,.
\end{equation}
Note that for the point mass~\eqref{eqn:pointmass}, \(U\) reduces to the previously introduced spherically symmetric Newtonian potential~\eqref{eqn:sphernewtpot}. Analogously, equation~\eqref{eqn:h200} governing \(h^{(2)}_{00}\) reduces to
\begin{equation}
h^{(2)}_{00,ii} = -c_2\rho\,,
\end{equation}
where \(c_2\) is given by equation~\eqref{eqn:c2def}, and thus has the solution
\begin{equation}\label{eqn:h200general}
h^{(2)}_{00} = \frac{c_2}{4\pi}U\,.
\end{equation}
We can compare this result to the corresponding metric component~\eqref{eqn:g00} in the spherically symmetric case. From this we see that the effective Newtonian constant is given by
\begin{equation}
G = \frac{c_2}{8\pi}\,.
\end{equation}
Here we have dropped the subscript ``eff'' in order to indicate that \(G\) is now really a constant, in contrast to being an effective quantity, which depends on the distance between the gravitating mass source and the test mass.

We proceed by solving for the metric component \(h^{(2)}_{ij}\). The corresponding equation~\eqref{eqn:h2ij} reduces to
\begin{equation}
h^{(2)}_{ij,kk} = -c_4\rho\delta_{ij}\,,
\end{equation}
where \(c_4\) is defined in equation~\eqref{eqn:c4def}. This is again a Poisson equation, which is solved by
\begin{equation}\label{eqn:h2ijgeneral}
h^{(2)}_{ij} = \frac{c_4}{4\pi}U\delta_{ij}\,.
\end{equation}
With this result we have determined the metric at the second velocity order.

\subsection{Third velocity order}\label{subsec:ppn3}
We now come to the third velocity order metric component \(h^{(3)}_{0i}\), which is determined by equation~\eqref{eqn:r30i}. In the case of a massless scalar field we consider in this section this equation reduces to
\begin{equation}
h^{(3)}_{0i,jj} = \frac{1}{G_{4(0,0)}}\rho v_i - \frac{c_2}{8\pi}U_{,0i}\,.
\end{equation}
Note that in contrast to section~\ref{subsec:beta} this equation is not satisfied identically under the assumptions made in this section, since we have not assumed that \(h^{(3)}_{0i}\) and \(v_i\) vanish and \(U\) is time independent. We thus find that the solution is given by
\begin{equation}\label{eqn:h30igeneral}
h^{(3)}_{0i} = \frac{1}{16\pi}\left[\left(c_2 - \frac{4}{G_{4(0,0)}}\right)V_i - c_2W_i\right]\,,
\end{equation}
where we used the third order PPN potentials
\begin{equation}
V_i(t,\vec{x}) = \int d^3x'\frac{\rho(t,\vec{x}')v_i(t,\vec{x}')}{|\vec{x} - \vec{x}'|}\,,\quad
W_i(t,\vec{x}) = \int d^3x'\frac{\rho(t,\vec{x}')v_j(t,\vec{x}')(x_i - x_i')(x_j - x_j')}{|\vec{x} - \vec{x}'|^3}\,.
\end{equation}
This result determines the metric at the third velocity order.

\subsection{Fourth velocity order}\label{subsec:ppn4}
We finally come to the solution of the gravitational field equations at the fourth velocity order. This calculation is considerably simpler than the corresponding calculation in section~\ref{subsec:beta}, since from our restriction \(K_{(2,0)} = 0\) follows that \(q_{16} = 0\), so that \(\psi^{(4)}\) does not appear in the fourth order metric equation~\eqref{eqn:h400}. We thus do not need to calculate \(\psi^{(4)}\) and can directly proceed with solving for \(h^{(4)}_{00}\). Inserting the solutions found for the second and third velocity order in sections~\ref{subsec:ppn2} and~\ref{subsec:ppn3}, equation~\eqref{eqn:h400} reduces to
\begin{equation}\label{eqn:h400ii}
h^{(4)}_{00,ii} = w_1U_{,00} + w_2U_{,i}U_{,i} + w_3\rho U + w_4\rho\Pi + w_5\rho v^2 + w_6p\,,
\end{equation}
where the constants \(w_1, \ldots, w_6\) take the values~\eqref{eqn:h400iicoeff} listed in appendix~\ref{sec:expcoeff}. The solution is given by
\begin{equation}\label{eqn:h400general}
h^{(4)}_{00} = \frac{w_2}{2}U^2 + \left(\frac{w_1}{2} - \frac{w_5}{4\pi}\right)\Phi_1 - \left(w_2 + \frac{w_3}{4\pi}\right)\Phi_2 - \frac{w_4}{4\pi}\Phi_3 - \frac{w_6}{4\pi}\Phi_4 - \frac{w_1}{2}\mathcal{A} - \frac{w_1}{2}\mathcal{B}\,,
\end{equation}
where the newly introduced PPN potentials are given by
\begin{gather}
\Phi_1(t,\vec{x}) = \int d^3x'\frac{\rho(t,\vec{x}')v^2(t,\vec{x}')}{|\vec{x} - \vec{x}'|}\,,\quad
\Phi_2(t,\vec{x}) = \int d^3x'\frac{\rho(t,\vec{x}')U(t,\vec{x}')}{|\vec{x} - \vec{x}'|}\,,\nonumber\\
\Phi_3(t,\vec{x}) = \int d^3x'\frac{\rho(t,\vec{x}')\Pi(t,\vec{x}')}{|\vec{x} - \vec{x}'|}\,,\quad
\Phi_4(t,\vec{x}) = \int d^3x'\frac{p(t,\vec{x}')}{|\vec{x} - \vec{x}'|}\,,\\
\mathcal{A}(t,\vec{x}) = \int d^3x'\frac{\rho(t,\vec{x}')\left[v_i(t,\vec{x}')(x_i - x_i')\right]^2}{|\vec{x} - \vec{x}'|^3}\,,\quad
\mathcal{B}(t,\vec{x}) = \int d^3x'\frac{\rho(t,\vec{x}')}{|\vec{x} - \vec{x}'|}(x_i - x_i')\frac{dv_i(t,\vec{x}')}{dt}\,.\nonumber
\end{gather}
With this result we have finally calculated all metric components to their respective velocity orders, which are required to determine the PPN parameters.

\subsection{PPN gauge and PPN parameters}\label{subsec:ppngauge}
We now use the solution for the metric components \(h^{(2)}_{00}\), \(h^{(2)}_{ij}\), \(h^{(3)}_{0i}\) and \(h^{(4)}_{00}\) calculated above in order to determine the PPN parameters. For this purpose, the metric must be in a particular gauge, in which it takes the form~\cite{Will:1993ns}
\begin{subequations}\label{eqn:standardppn}
\begin{align}
\bar{h}^{(2)}_{\bar{0}\bar{0}} &= 2U\,,\\
\bar{h}^{(2)}_{\bar{i}\bar{j}} &= 2\gamma U\delta_{ij}\,,\\
\bar{h}^{(3)}_{\bar{0}\bar{i}} &= -\frac{1}{2}(3 + 4\gamma + \alpha_1 - \alpha_2 + \zeta_1 - 2\xi)V_i - \frac{1}{2}(1 + \alpha_2 - \zeta_1 + 2\xi)W_i\,,\\
\bar{h}^{(4)}_{\bar{0}\bar{0}} &= -2\beta U^2 - 2\xi\Phi_W + (2 + 2\gamma + \alpha_3 + \zeta_1 - 2\xi)\Phi_1 + 2(1 + 3\gamma - 2\beta + \zeta_2 + \xi)\Phi_2\\
&\phantom{=}+ 2(1 + \zeta_3)\Phi_3 + 2(3\gamma + 3\zeta_4 - 2\xi)\Phi_4 - (\zeta_1 - 2\xi)\mathcal{A}\,,\nonumber
\end{align}
\end{subequations}
where in addition to the previously listed PPN potentials we have also included the Whitehead term
\begin{equation}
\Phi_W(t,\vec{x}) = \int d^3x'd^3x''\rho(t,\vec{x}')\rho(t,\vec{x}'')\frac{x_i - x_i'}{|\vec{x} - \vec{x}'|^3}\left(\frac{x_i' - x_i''}{|\vec{x} - \vec{x}''|} - \frac{x_i - x_i''}{|\vec{x}' - \vec{x}''|}\right)\,.
\end{equation}
Note in particular that this metric must not contain the PPN potential \(\mathcal{B}\), in contrast to our result~\eqref{eqn:h400general}. This indicates that the solution we found is not yet in the PPN gauge. We thus need to eliminate the potential \(\mathcal{B}\) by a suitable gauge transformation, i.e., by a suitable change of coordinates. It turns out that the PPN gauge is achieved by introducing new coordinates \(\bar{x}^{\bar{\mu}}\) given by
\begin{equation}
\bar{x}^{\bar{\mu}} = x^{\mu} + \xi^{\mu}\,, \quad \xi_0 = -\frac{w_1}{4}\chi_{,0}\,, \quad \xi_i = 0\,,
\end{equation}
where \(\chi\) is the superpotential
\begin{equation}
\chi(t,\vec{x}) = -\int d^3x'\rho(t,\vec{x}')|\vec{x} - \vec{x}'|\,.
\end{equation}
This gauge transformation changes the metric to
\begin{equation}
\bar{g}_{\bar{0}\bar{0}} = g_{00} + \frac{w_1}{2}(\mathcal{A} + \mathcal{B} - \Phi_1)\,, \quad \bar{g}_{\bar{0}\bar{i}} = g_{0i} + \frac{w_1}{4}(V_i - W_i)\,, \quad \bar{g}_{\bar{i}\bar{j}} = g_{ij}\,.
\end{equation}
In this new gauge we thus find the metric components
\begin{subequations}
\begin{align}
\bar{h}^{(2)}_{\bar{0}\bar{0}} &= \frac{c_2}{4\pi}U\,,\\
\bar{h}^{(2)}_{\bar{i}\bar{j}} &= \frac{c_4}{4\pi}U\delta_{ij}\,,\\
\bar{h}^{(3)}_{\bar{0}\bar{i}} &= \frac{1}{16\pi}\left[\left(c_2 - \frac{4}{G_{4(0,0)}} + \frac{w_1}{4}\right)V_i - \left(c_2 + \frac{w_1}{4}\right)W_i\right]\,,\\
\bar{h}^{(4)}_{\bar{0}\bar{0}} &= \frac{w_2}{2}U^2 - \frac{w_5}{4\pi}\Phi_1 - \left(w_2 + \frac{w_3}{4\pi}\right)\Phi_2 - \frac{w_4}{4\pi}\Phi_3 - \frac{w_6}{4\pi}\Phi_4\,,
\end{align}
\end{subequations}
up to the required velocity orders. It is conventional to work in the normalization \(G \equiv 1\), in which one can directly read off the PPN parameters from the metric equation~\eqref{eqn:standardppn}. This normalization is obtained by multiplying the gravitational part of the action~\eqref{eqn:action}, and thus the free functions \(K, G_3, G_4, G_5\), with the constant \(c_2/8\pi\). After applying this normalization we find the metric components
\begin{subequations}
\begin{align}
\bar{h}^{(2)}_{\bar{0}\bar{0}} &= 2U\,,\\
\bar{h}^{(2)}_{\bar{i}\bar{j}} &= 2\frac{c_4}{c_2}U\delta_{ij}\,,\\
\bar{h}^{(3)}_{\bar{0}\bar{i}} &= \left(\frac{1}{2} - \frac{2}{G_{4(0,0)}c_2} + \frac{w_1}{8c_2}\right)V_i - \left(\frac{1}{2} + \frac{w_1}{8c_2}\right)W_i\,,\\
\bar{h}^{(4)}_{\bar{0}\bar{0}} &= 32\pi^2\frac{w_2}{c_2^2}U^2 - 2\frac{w_5}{c_2}\Phi_1 - \frac{64\pi^2}{c_2^2}\left(w_2 + \frac{w_3}{4\pi}\right)\Phi_2 - 2\frac{w_4}{c_2}\Phi_3 - 2\frac{w_6}{c_2}\Phi_4\,.
\end{align}
\end{subequations}
We can now compare this to the standard form~\eqref{eqn:standardppn} of the PPN metric. Reading off the PPN parameters and inserting the previously introduced constants listed in appendix~\ref{sec:expcoeff} we finally obtain
\begin{equation}\label{eqn:mlfullppn}
\gamma = \frac{\omega + 1}{\omega + 2}\,, \quad \beta = 1 + \frac{\omega + \tau - 4\omega\sigma}{4(\omega + 2)^2(2\omega + 3)}\,, \quad \alpha_1 = \alpha_2 = \alpha_3 = \zeta_1 = \zeta_2 = \zeta_3 = \zeta_4 = \xi = 0\,,
\end{equation}
where we used the abbreviations~\eqref{eqn:omega} and~\eqref{eqn:abbreviations}. This result shows that Horndeski's theory of gravity belongs to the class of fully conservative theories, in which momentum and angular momentum are conserved, and in which there are no preferred-frame effects. Theories of this type are characterized by the PPN parameters \(\alpha_1 = \alpha_2 = \alpha_3 = \zeta_1 = \zeta_2 = \zeta_3 = \zeta_4 = 0\). Further, it is free of preferred-location effects, or Whitehead effects, which is indicated by the vanishing Whitehead parameter \(\xi\). This leaves only the two PPN parameters \(\gamma\) and \(\beta\) which potentially deviate from observations, as we will argue in the following section.

\section{Comparison with observations}\label{sec:obs}
We now briefly compare the results obtained in the previous two sections to the values of the PPN parameters measured in solar system experiments. We restrict our discussion to the PPN parameters \(\gamma\) and \(\beta\), since for a massive scalar field these are the only parameters we have calculated in section~\ref{sec:sphersym}, while for a massless scalar field these are the only non-trivial PPN parameters, according to our calculation in section~\ref{sec:mlfullppn}, where we have seen that all other parameters take the value \(0\), in agreement with observations independently of the choice of a particular theory from Horndeski's class. See~\cite{Will:2014xja} for a recent review of the values of the full set of PPN parameters.

We start our discussion with the case of a massive scalar field considered in section~\ref{sec:sphersym}. Here we can essentially distinguish two regimes: a light scalar field / short interaction distance with \(m_{\psi}r \ll 1\), and a heavy scalar field / long interaction distance with \(m_{\psi}r \gg 1\). We will not consider the intermediate case, as it can simply be obtained by interpolation. In the limit of a light scalar field, the values of the PPN parameters approach their values in the massless case, which we will discuss later in this section, so that for now we will focus on the heavy scalar field case. Note that the PPN parameters \(\gamma\) and \(\beta\) shown in~\eqref{eqn:gamma} and~\eqref{eqn:beta} depend exponentially on \(m_{\psi}r\). Keeping only the leading order terms we find
\begin{equation}
\gamma = 1 - \frac{2}{2\omega + 3}e^{-m_{\psi}r} + \mathcal{O}(e^{-2m_{\psi}r})\,, \quad \beta = 1 + \frac{m_{\psi}r}{2\omega + 3}\ln(m_{\psi}r)e^{-m_{\psi}r} + \mathcal{O}(e^{-2m_{\psi}r})\,,
\end{equation}
so that the PPN parameters in this limiting case depend only on the constants \(m_{\psi}\) and \(\omega\). In order to derive bounds on these constants, we must consider measurements of \(\gamma\) and \(\beta\) at a fixed interaction distance \(r = r_0\). Currently the most stringent bounds of this type are obtained from the time delay of radar signals sent between Earth and the Cassini spacecraft on its way to Saturn~\cite{Bertotti:2003rm}. The experiment yielded the value \(\gamma - 1 = (2.1 \pm 2.3) \cdot 10^{-5}\). The radio signals were passing by the Sun at a distance of \(1.6\) solar radii or \(r_0 \approx 7.44 \cdot 10^{-3}\mathrm{AU}\). The excluded parameter region obtained from this experiment has already been derived in a previous work; see~\cite{Hohmann:2013rba} for a full discussion.

The second case we consider is that of a massless (or light) scalar field as discussed in section~\ref{sec:mlfullppn}, for which the PPN parameters approach the values~\eqref{eqn:mlfullppn}, and are thus independent of the interaction distance. We may therefore also use bounds on the PPN parameters from experiments for which an interaction distance cannot be easily defined, such as the latest ephemeris releases INPOP13~\cite{Verma:2013ata,Fienga:2014,Fienga:2014bvy}. The bounds obtained from these datasets are given by \(\gamma - 1 = (-0.3 \pm 2.5) \cdot 10^{-5}\) and \(\beta - 1 = (0.2 \pm 2.5) \cdot 10^{-5}\). However, it turns out that the Cassini bound on \(\gamma\), and thus on \(\omega\), is still more stringent, and yields \(\omega \geq 4.0 \cdot 10^4\) at \(2\sigma\) confidence level. From the bound on \(\beta\) we then obtain the most stringent bound \(-2.5 \cdot 10^{10} \leq \tau - 4\omega\sigma \leq 2.7 \cdot 10^{10}\) at \(2\sigma\) confidence level for \(\omega = 4.0 \cdot 10^4\), and less stringent bounds for larger values of \(\omega\).

This concludes our discussion of the post-Newtonian limit for the general class of Horndeski's gravity theories. The result we obtained can now easily be applied to particular theories within this class. We will show three examples in the following section.

\section{Examples}\label{sec:examples}
After discussing the post-Newtonian limit of the most general form of Horndeski's theory compatible with our assumptions in the previous sections, we now come to particular example theories which fall into this class. In the following we will derive only the relevant PPN parameters \(\gamma\) and \(\beta\) for these theories, as we argued in the preceding section. In particular, we will discuss a common class of scalar-tensor theories with arbitrary potential in section~\ref{subsec:stgpot}, generalized Higgs inflation in section~\ref{subsec:genhiggs} and the galileon model in section~\ref{subsec:galileons}.

\subsection{Scalar-tensor gravity with a general potential}\label{subsec:stgpot}
As we already mentioned in the introductory section~\ref{sec:motivation}, the work presented in this article generalizes a result obtained for a more restricted class of scalar-tensor theories~\cite{Hohmann:2013rba}. The starting point of this earlier work is an action of the form
\begin{equation}
S = \frac{1}{2\kappa^2}\int d^4x\sqrt{-g}\left(\phi R - \frac{\omega(\phi)}{\phi}\partial_{\rho}\phi\partial^{\rho}\phi - 2\kappa^2V(\phi)\right) + S_m[g_{\mu\nu},\chi_m]\,,
\end{equation}
where the potential \(V(\phi)\) and the kinetic coupling function \(\omega(\phi)\) are free functions of the scalar field. Note that this action is written in the so-called Jordan frame, which is most convenient for calculating the PPN parameters and which can directly be compared to the action~\eqref{eqn:action} we use in this article. Note, however, that one can also write this action in the so-called Einstein frame by application of a conformal transformation, and that the PPN parameters can also be calculated in this frame~\cite{Scharer:2014kya}. This invariance of the theory under conformal transformations also allows expressing its PPN parameters in terms of invariants under these transformations~\cite{Jarv:2014hma}.

Since we use the action in the Jordan frame, we can directly compare it to the Horndeski gravity action~\eqref{eqn:action} and read off the functions
\begin{equation}
K(\phi,X) = \frac{\omega(\phi)}{\kappa^2\phi}X - V(\phi)\,, \quad G_4(\phi,X) = \frac{\phi}{2\kappa^2}\,, \quad G_3(\phi,X) = G_5(\phi,X) = 0\,.
\end{equation}
We then expand these functions in a Taylor series around the cosmological background value \(\phi = \Phi\). The relevant, non-vanishing coefficients in this Taylor series are given by
\begin{gather}
K_{(0,0)} = -V_0\,, \quad K_{(1,0)} = -V_1\,, \quad K_{(2,0)} = -V_2\,, \quad  K_{(3,0)} = -V_3\,,\nonumber\\
K_{(0,1)} = \frac{\omega_0}{\kappa^2\Phi}\,, \quad K_{(1,1)} = \frac{\Phi\omega_1 - \omega_0}{\kappa^2\Phi^2}\,, \quad G_{4(0,0)} = \frac{\Phi}{2\kappa^2}\,, \quad G_{4(1,0)} = \frac{1}{2\kappa^2}\,.
\end{gather}
Here we have expanded the functions \(\omega(\phi)\) and \(V(\phi)\) in analogy to the expansion~\eqref{eqn:taylorseries}, which takes the form
\begin{equation}
\omega(\phi) = \omega_0 + \omega_1\psi + \mathcal{O}(\psi^2)\,,\quad
V(\phi) = V_0 + V_1\psi + V_2\psi^2 + V_3\psi^3 + \mathcal{O}(\psi^4)\,,
\end{equation}
where \(\phi = \Phi + \psi\). The constraint~\eqref{eqn:bgcondition}, which ensures the validity of the post-Newtonian approximation and its consistency with the cosmological background, takes the form \(V_0 = V_1 = 0\), while the constraint~\eqref{eqn:squarerestrict} is satisfied identically. For the constants determining the PPN parameters defined in equations~\eqref{eqn:mpsi},~\eqref{eqn:omega} and~\eqref{eqn:abbreviations} we then find the values
\begin{equation}
m_{\psi} = 2\kappa\sqrt{\frac{V_2\Phi}{2\omega_0 + 3}}\,, \quad \omega = \omega_0\,, \quad \tau = \omega_1\Phi - \omega_0\,, \quad \sigma = 0\,, \quad \mu = -2\kappa^2\Phi^2V_3\,.
\end{equation}
Finally, we make use of these values in order to obtain the PPN parameters. These are given by
\begin{equation}
\gamma(r) = \frac{2\omega_0 + 3 - e^{-m_{\psi}r}}{2\omega_0 + 3 + e^{-m_{\psi}r}}
\end{equation}
and
\begin{equation}
\begin{split}
\beta(r) &= 1 + \frac{1}{(2\omega_0 + 3 + e^{-m_{\psi}r})^2}\Bigg\{\frac{\Phi\omega_1}{2\omega_0 + 3}e^{-2m_{\psi}r}\\
&\phantom{=}+ (2\omega_0 + 3)m_{\psi}r\left[e^{-m_{\psi}r}\ln(m_{\psi}r) - \left(m_{\psi}r + e^{m_{\psi}r}\right)\mathrm{Ei}(-2m_{\psi}r) - \frac{1}{2}e^{-2m_{\psi}r}\right]\\
&\phantom{=}+ \frac{3}{2}\left(1 - \frac{\Phi V_3}{V_2} + \frac{\Phi\omega_1}{2\omega_0 + 3}\right)m_{\psi}r\left[e^{m_{\psi}r}\mathrm{Ei}(-3m_{\psi}r) - e^{-m_{\psi}r}\mathrm{Ei}(-m_{\psi}r)\right]\Bigg\}\,,
\end{split}
\end{equation}
which reproduces our previously found result. Finally, the massless case is given by \(V_2 = V_3 = 0\) and yields the PPN parameters
\begin{equation}
\gamma = \frac{\omega_0 + 1}{\omega_0 + 2}\,, \quad \beta = 1 + \frac{\Phi\omega_1}{(2\omega_0 + 3)(2\omega_0 + 4)^2}\,,
\end{equation}
which is a well-known result for the PPN parameters of a massless scalar-tensor theory~\cite{Nordtvedt:1970uv}.

\subsection{Generalized Higgs inflation}\label{subsec:genhiggs}
We now discuss a class of models whose basic idea is to identify the Higgs field with the inflaton, which is the scalar field responsible for the inflation in the early universe, and which can thus be summarized under the name Higgs inflation models. It has been shown that a number of these models can nicely be written as a particular subclass of Horndeski's gravity theory, which has been called generalized Higgs inflation~\cite{Kamada:2012se}. The functions in the action~\eqref{eqn:action} of this model take the form
\begin{gather}
K(\phi,X) = \mathcal{K}(\phi)X - V(\phi)\,,\quad
G_3(\phi,X) = h_3(\phi)X\,,\nonumber\\
G_4(\phi,X) = g(\phi) + h_4(\phi)X\,,\quad
G_5(\phi,X) = h_5(\phi)X
\end{gather}
with six free functions \(\mathcal{K}, V, g, h_3, h_4, h_5\) of the scalar field \(\phi\). After performing a Taylor expansion of these functions in analogy to the expansion~\eqref{eqn:taylorseries} we find the relevant, non-vanishing Taylor coefficients
\begin{gather}
K_{(0,0)} = -V_0\,, \quad K_{(1,0)} = -V_1\,, \quad K_{(2,0)} = -V_2\,, \quad K_{(3,0)} = -V_3\,, \quad K_{(0,1)} = \mathcal{K}_0\,, \quad K_{(1,1)} = \mathcal{K}_1\,,\nonumber\\
G_{3(0,1)} = h_{3,0}\,, \quad G_{4(0,0)} = g_0\,, \quad G_{4(1,0)} = g_1\,, \quad G_{4(2,0)} = g_2\,, \quad G_{4(0,1)} = h_{4,0}\,, \quad G_{4(1,1)} = h_{4,1}\,.\label{eqn:genhiggscoeffs}
\end{gather}
Similarly to the scalar-tensor theory discussed in the previous section we find that the constraint~\eqref{eqn:bgcondition}, which ensures the validity of our perturbative expansion around a stationary cosmological background, takes the form \(V_0 = V_1 = 0\). Further, the constraint~\eqref{eqn:squarerestrict} translates to \(h_{3,0} = 3h_{4,1}\) and \(h_{4,0} = 0\), which we will impose in the remainder of this section. We then find the mass of the scalar field and the constants
\begin{equation}\label{eqn:genhiggscons}
m_{\psi} = \sqrt{\frac{2g_0V_2}{g_0\mathcal{K}_0 + 3g_1^2}}\,, \quad \omega = \frac{g_0\mathcal{K}_0}{2g_1^2}\,, \quad \sigma = \frac{g_0g_2}{g_1^2}\,, \quad \tau = \frac{g_0^2\mathcal{K}_1}{2g_1^3}\,, \quad \mu = -\frac{g_0^2V_3}{g_1^3}\,,
\end{equation}
from which the PPN parameters \(\gamma\) and \(\beta\) can be obtained.

We discuss a few special cases listed in~\cite{Kamada:2012se}, which can be viewed as corrections to general relativity with a minimally coupled scalar field given by the gravitational action
\begin{equation}
S_G = \int d^4x\sqrt{-g}\left(\frac{M_{\text{Pl}}^2}{2}R + X - V(\phi) + \Delta\mathcal{L}\right)\,,
\end{equation}
where \(M_{\text{Pl}}\) is the Planck mass. Without any such correction \(\Delta\mathcal{L}\) the only non-vanishing Taylor coefficients are \(g_0 = M_{\text{Pl}}^2/2\), \(\mathcal{K}_0 = 1\) and the terms originating from the potential \(V(\phi)\). In particular, we consider the following models:

\begin{itemize}
\item
Running kinetic inflation~\cite{Nakayama:2010kt,Nakayama:2010sk}: \(\Delta\mathcal{L} = \kappa\phi^{2n}X\) with parameters \(\kappa\) and \(n\). In this model we obtain the modified Taylor coefficients \(\mathcal{K}_0 = 1 + \kappa\Phi^{2n}\) and \(\mathcal{K}_1 = 2n\kappa\Phi^{2n - 1}\). In this case we have \(g_1 = 0\), so that we obtain the limit \(\omega \to \infty\), from which follows \(\gamma = \beta = 1\).

\item
Higgs G-inflation~\cite{Kamada:2010qe}: \(\Delta\mathcal{L} = -\phi X\square\phi/M^4\) with parameter \(M\). In this model the only modified Taylor coefficient is \(h_{3,0} = -\Phi/M^4\). However, this coefficient is restricted by the condition~\eqref{eqn:squarerestrict}, so that this model will yield terms in the gravitational field equations which are not covered by the PPN formalism we used in this article.

\item
Non-minimal Higgs inflation:~\cite{Spokoiny:1984bd,Futamase:1987ua,Bezrukov:2007ep}: \(\Delta\mathcal{L} = -\xi\phi^2R/2\) with parameter \(\xi\). In this model we find the modified Taylor coefficients
\begin{equation}
g_0 = \frac{M_{\text{Pl}}^2 - \xi\Phi^2}{2}\,, \quad g_1 = -\xi\Phi\,, \quad g_2 = -\frac{\xi}{2}\,,
\end{equation}
from which follow the constants
\begin{equation}
m_{\psi} = M_{\text{eff}}\sqrt{\frac{2V_2}{M_{\text{eff}}^2 + 6\xi^2\Phi^2}}\,, \quad \omega = \frac{M_{\text{eff}}^2}{4\xi^2\Phi^2}\,, \quad \sigma = -\frac{M_{\text{eff}}^2}{4\xi\Phi^2}\,, \quad \tau = 0\,, \quad \mu = \frac{M_{\text{eff}}^4V_3}{4\xi^3\Phi^3}\,,
\end{equation}
where we introduced the effective Planck mass \(M_{\text{eff}}^2 = M_{\text{Pl}}^2 - \xi\Phi^2\). Typically one is interested in the case \(M_{\text{eff}} \approx M_{\text{Pl}}\). Here we make use of the fact that \(M_{\text{Pl}} \gg \Phi \approx 246\mathrm{GeV}\), which is the vacuum expectation value of the Higgs field, and expand the PPN parameters \(\gamma\) and \(\beta\) in orders of \(\Phi/M_{\text{eff}}\). Up to the first non-trivial order we find
\begin{equation}
\gamma = 1 - 4\xi^2e^{-m_{\psi}r}\frac{\Phi^2}{M_{\text{eff}}^2} + \mathcal{O}\left(\frac{\Phi^3}{M_{\text{eff}}^3}\right)
\end{equation}
and
\begin{multline}
\beta = 1 + \big\{2\xi^3e^{-2m_{\psi}r} - \xi^2m_{\psi}r\big[e^{-2m_{\psi}r} - 2e^{-m_{\psi}r}\ln(m_{\psi}r)\\
+ 2 (m_{\psi}r + e^{m_{\psi}r})\mathrm{Ei}(-2m_{\psi}r)\big]\big\}\frac{\Phi^2}{M_{\text{eff}}^2} + \mathcal{O}\left(\frac{\Phi^3}{M_{\text{eff}}^3}\right)\,.
\end{multline}

\item
New Higgs inflation~\cite{Germani:2010gm,Germani:2010ux,Granda:2010hb,Granda:2011zk}:
\begin{equation}
\Delta\mathcal{L} = \frac{1}{2\tilde{\mu}^2}\left[XR + (\square\phi)^2 - (\nabla_{\mu}\nabla_{\nu}\phi)^2\right]
\end{equation}
with parameter \(\tilde{\mu}\). In this model the only modified Taylor coefficient is \(h_{4,0} = 1/2\tilde{\mu}^2\). However, as it was also the case for Higgs G-inflation, this coefficient is restricted by the condition~\eqref{eqn:squarerestrict}, so that also this model will yield terms in the gravitational field equations which are not covered by the PPN formalism we used in this article.

\item
Running Einstein inflation~\cite{Kamada:2012se}:
\begin{equation}
\Delta\mathcal{L} = \frac{\phi}{\Lambda^6}\left[XG_{\mu\nu}\nabla^{\mu}\nabla^{\nu}\phi - \frac{1}{6}(\square\phi)^3 + \frac{1}{2}(\square\phi)(\nabla_{\mu}\nabla_{\nu}\phi)^2 - \frac{1}{3}(\nabla_{\mu}\nabla_{\nu}\phi)^3\right]
\end{equation}
with parameter \(\Lambda\). This term does not influence any relevant Taylor coefficients, so that again we find \(\gamma = \beta = 1\).
\end{itemize}

We thus see that the non-minimal Higgs inflation model is the only model from which we obtain PPN parameters which potentially deviate from observations. Note, however, that for a scenario in which \(m_{\psi} \approx 125\mathrm{GeV}\) is the Higgs mass, any observable gravitational interaction takes place in the limit \(m_{\psi}r \gg 1\), such that one has the limiting values \(\gamma = \beta = 1\).

\subsection{Galileons}\label{subsec:galileons}
The last example we consider here is a scalar field whose action, in flat spacetime, is invariant under Galilean transformations, and hence is called galileon~\cite{Nicolis:2008in}. Here we consider the covariant theory in curved spacetime~\cite{Deffayet:2009wt}. The action is given by
\begin{equation}
\begin{split}
S = \bigintsss d^4x\sqrt{-g}\Bigg\{R &+ C_4\nabla_{\lambda}\phi\nabla^{\lambda}\phi\left[2(\square\phi)^2 - 2(\nabla_{\mu}\nabla_{\nu}\phi)^2 - \frac{1}{2}\nabla_{\mu}\phi\nabla^{\mu}\phi R\right]\\
&+ \frac{5}{2}C_5\nabla_{\lambda}\phi\nabla^{\lambda}\phi\big[(\square\phi)^3 - 3(\square\phi)(\nabla_{\mu}\nabla_{\nu}\phi)^2 + 2(\nabla_{\mu}\nabla_{\nu}\phi)^3\\
&- 6G_{\nu\rho}\nabla_{\mu}\phi\nabla^{\mu}\nabla^{\nu}\phi\nabla^{\rho}\phi\big] + C_1\phi + C_2\nabla_{\mu}\phi\nabla^{\mu}\phi + C_3\nabla_{\mu}\phi\nabla^{\mu}\phi\square\phi\Bigg\}\,.
\end{split}
\end{equation}
After integration by parts one can see that this action has the form of the Horndeski action~\eqref{eqn:action}, where the free functions are given by
\begin{equation}
K(\phi,X) = C_1\phi - 2C_2X\,, \quad G_3(\phi,X) = 2C_3X\,, \quad G_4(\phi,X) = 1 - 2C_4X^2\,, \quad G_5(\phi,X) = 15C_5X^2\,.
\end{equation}
The relevant, non-vanishing Taylor coefficients are thus given by
\begin{equation}
K_{(0,0)} = C_1\Phi\,, \quad K_{(1,0)} = C_1\,, \quad K_{(0,1)} = -2C_2\,, \quad G_{3(0,1)} = 2C_3\,, \quad G_{4(0,0)} = 1\,.
\end{equation}
We see that the terms involving \(C_4\) and \(C_5\) do not enter the relevant Taylor coefficients, and so have no influence on the post-Newtonian limit. The conditions~\eqref{eqn:bgcondition} and~\eqref{eqn:squarerestrict}, which we imposed in order for the post-Newtonian limit to be valid, require that \(C_1 = C_3 = 0\). From \(K_{(2,0)} = K_{(3,0)} = 0\) further follows that the scalar field is massless and we can apply the formalism detailed in section~\ref{sec:mlfullppn}. It turns out that \(\gamma = \beta = 1\) for this class of theories.

This concludes our discussion of particular examples for Horndeski gravity theories. We have seen that among this class there are several theories whose post-Newtonian limit is consistent with the solar system observations displayed in section~\ref{sec:obs}, as they are in particular compatible with the values \(\gamma = \beta = 1\) obtained for various example theories.

\section{Conclusion}\label{sec:conclusion}
In this article we discussed the post-Newtonian limit of Horndeski's theory of gravity. We showed that the post-Newtonian limit is fully determined by fifteen constant parameters, which arise as coefficients in the Taylor expansion of the free functions \(K, G_3, G_4, G_5\) in the Horndeski Lagrangian around the cosmological background value of the scalar field. It turned out that for the post-Newtonian limit to be consistent, we must impose several constraints on these Taylor coefficients. With these constraints in place, we calculated the post-Newtonian limit in two different scenarios.

In the first scenario we considered the most general theory consistent with the aforementioned constraints. We defined the PPN parameters \(\gamma\) and \(\beta\) and calculated their values for a static point mass. It turned out that the PPN parameters are not constant, but depend on the distance between the gravitating source and the test mass.

In the second scenario we imposed additional constraints on the Taylor coefficients under which the scalar field becomes massless. We showed that when these constraints are satisfied, the post-Newtonian limit of the theory assumes the standard PPN form, which is characterized by ten constant PPN parameters. We calculated these parameters and showed that Horndeski's theory is a fully conservative theory, which is free of preferred-frame and preferred-location effects, which means that only the parameters \(\gamma\) and \(\beta\) potentially deviate from their observed values.

We finally applied our analysis to a number of example theories, including a previously discussed scalar-tensor theory with arbitrary scalar potential, various Higgs inflation models and galileons.

The work presented here allows for further extensions and generalizations. A straightforward generalization is to drop the assumption that the cosmological background value \(\Phi\) of the scalar field is constant and to allow for a time dependence \(\dot{\Phi} \neq 0\). Another possibility is to investigate the parameterized post-Newtonian limit of more general scalar-tensor theories beyond the Horndeski Lagrangian, which introduce higher order derivatives into the gravitational field equations. Despite originally being regarded as ill-defined due to Ostrogradski instabilities and ghosts, it has turned out that these problems may be overcome and healthy theories exist~\cite{Zumalacarregui:2013pma,Gleyzes:2014dya,Gleyzes:2014qga,Gao:2014soa,Fasiello:2014aqa,Gao:2014fra,DeFelice:2015isa}. Yet another possible direction of future research is to consider theories with several scalar degrees of freedom, such as the recently developed generalization of Horndeski's theory to two scalar fields~\cite{Ohashi:2015fma}. Finally, one may also consider modifications of the formalism itself, in order to include effects caused by screening mechanisms such as the Vainshtein mechanism~\cite{Avilez-Lopez:2015dja}.

\appendix

\section{PPN expansion coefficients}\label{sec:expcoeff}
This appendix lists the coefficients appearing in several lengthy equations and intermediate results of the calculation of the PPN parameters in sections~\ref{sec:sphersym} and~\ref{sec:mlfullppn}. The following coefficients appear in the scalar field equation~\eqref{eqn:psi4} at the fourth velocity order:
\begin{gather}
d_3 = -\frac{(K_{(1,1)} - 4G_{3(2,0)})G_{4(0,0)} + (K_{(0,1)} - 2G_{3(1,0)} + 12G_{4(2,0)})G_{4(1,0)}}{2(K_{(0,1)} - 2G_{3(1,0)})G_{4(0,0)} + 6G_{4(1,0)}^2}\,,\nonumber\\
d_2 = \frac{2K_{(2,0)}G_{4(1,0)} - 3K_{(3,0)}G_{4(0,0)}}{(K_{(0,1)} - 2G_{3(1,0)})G_{4(0,0)} + 3G_{4(1,0)}^2}\,,\quad
d_4 = -\frac{(K_{(1,1)} - 4G_{3(2,0)})G_{4(0,0)} + 6G_{4(1,0)}G_{4(2,0)}}{(K_{(0,1)} - 2G_{3(1,0)})G_{4(0,0)} + 3G_{4(1,0)}^2}\,,\nonumber\\
d_5 = -d_6 = \frac{(G_{3(0,1)} - 3G_{4(1,1)})G_{4(0,0)} - (G_{4(0,1)} - G_{5(1,0)})G_{4(1,0)}}{(K_{(0,1)} - 2G_{3(1,0)})G_{4(0,0)} + 3G_{4(1,0)}^2}\,,\label{eqn:psi4coeff}\\
d_8 = \frac{G_{4(1,0)}^2 - G_{4(0,0)}G_{4(2,0)}}{(K_{(0,1)} - 2G_{3(1,0)})G_{4(0,0)} + 3G_{4(1,0)}^2}\,,\quad
d_{10} = \frac{2(G_{4(0,1)} - G_{5(1,0)})G_{4(0,0)}}{(K_{(0,1)} - 2G_{3(1,0)})G_{4(0,0)} + 3G_{4(1,0)}^2}\,,\nonumber\\
d_{12} = -3d_{11} = \frac{3G_{4(1,0)}}{2(K_{(0,1)} - 2G_{3(1,0)})G_{4(0,0)} + 6G_{4(1,0)}^2}\,,\quad
d_1 = d_7 = d_9 = 1\,.\nonumber
\end{gather}
From this equation one derives equation~\eqref{eqn:psi42}, which contains the following coefficients:
\begin{gather}
e_2 = \frac{M^2c_{\psi}^2}{32\pi^2}\left[2d_2 + (d_7 + d_8)c_1 - (d_7 + 3d_8 - 2d_9)c_3 + 2(d_3 + d_4)m_{\psi}^2 + 2d_8m_{\psi}^2\frac{G_{4(1,0)}}{G_{4(0,0)}}\right]\,,\nonumber\\
e_3 = e_5m_{\psi} = \frac{M^2c_{\psi}}{32\pi^2m_{\psi}}d_7\left[(c_3 - c_1)c_{\psi} + (c_2 - c_4)m_{\psi}^2\right]\,,\label{eqn:psi42coeff}\\
e_1 = \frac{M^2c_{\psi}}{16\pi^2}d_9\left(c_4m_{\psi}^2 - c_3c_{\psi}\right)\,,\quad
e_4 = e_6m_{\psi} = \frac{M^2c_{\psi}^2}{32\pi^2m_{\psi}^2}\left[d_7(c_1 - c_3) + 2d_3m_{\psi}^2\right]\,.\nonumber
\end{gather}
In terms of these coefficients we express the coefficients of the solution~\eqref{eqn:psi4sol} in the form
\begin{gather}
f_4 = -\frac{e_2}{2m_{\psi}} + \frac{e_4}{2} - \frac{e_6m_{\psi}}{4}\,,\quad
f_6 = \frac{e_2}{2m_{\psi}} - \frac{3e_4}{2} + \frac{9e_6m_{\psi}}{4}\,,\nonumber\\
f_1 = \frac{e_5}{2}\,,\quad
f_2 = \frac{e_6}{2}\,,\quad
f_3 = -\frac{e_1}{2m_{\psi}}\,,\quad
f_5 = \frac{e_1}{2m_{\psi}} - e_3 + e_5m_{\psi}\,.\label{eqn:psi4solcoeff}
\end{gather}
A similar expansion is used for the metric component \(h^{(4)}_{00}\). The coefficients in equation~\eqref{eqn:h400} are listed below:
\begin{gather}
q_1 = \frac{G_{4(1,0)}}{G_{4(0,0)}}(d_1 - 1)\,,\quad
q_2 = \frac{G_{4(1,0)}d_2 - K_{(2,0)}}{G_{4(0,0)}}\,,\quad
q_3 = \frac{G_{4(1,0)}d_3 + 2G_{4(2,0)}}{G_{4(0,0)}}\,,\nonumber\\
q_4 = \frac{G_{4(1,0)}d_4 + 2G_{4(2,0)}}{G_{4(0,0)}}\,,\quad
q_5 = \frac{G_{4(1,0)}}{G_{4(0,0)}}d_5\,,\quad
q_6 = \frac{G_{4(1,0)}}{G_{4(0,0)}}d_6\,,\quad
q_7 = \frac{G_{4(1,0)}}{G_{4(0,0)}}(d_7 - 1)\,,\nonumber\\
q_8 = \frac{G_{4(1,0)}}{G_{4(0,0)}}d_8\,,\quad
q_9 = \frac{G_{4(1,0)}}{G_{4(0,0)}}(d_9 - 1)\,,\quad
q_{10} = \frac{G_{4(1,0)}}{G_{4(0,0)}}d_{10}\,,\quad
q_{11} = 0\,,\label{eqn:h400coeff}\\
q_{12} = -\frac{G_{4(1,0)}}{G_{4(0,0)}}\,,\quad
q_{13} = 2\frac{G_{4(1,0)}}{G_{4(0,0)}}\,,\quad
q_{14} = -1\,,\quad
q_{15} = 1\,,\quad
q_{16} = \frac{G_{4(1,0)}}{G_{4(0,0)}}m_{\psi}^2\,,\nonumber\\
q_{17} = \frac{2G_{4(1,0)}d_{11} - 1}{2G_{4(0,0)}}\,,\quad
q_{18} = -\frac{1}{G_{4(0,0)}}\,,\quad
q_{19} = \frac{1}{2G_{4(0,0)}}\,,\quad
q_{20} = \frac{2G_{4(1,0)}d_{12} - 3}{2G_{4(0,0)}}\,.\nonumber
\end{gather}
From this equation one obtains equation~\eqref{eqn:h4002}, which contains the following coefficients:
\begin{gather}
s_2 = q_{16}f_2 + \frac{M^2c_{\psi}^2}{32\pi^2}\Bigg[2q_2 + 2(q_3 + q_4)m_{\psi}^2 + q_7(c_3 - c_1) + q_8\left(c_1 - 3c_3 + 2\frac{G_{4(1,0)}}{G_{4(0,0)}}c_{\psi}\right)\nonumber\\
\phantom{=}+ 2q_9c_3 + (2q_{11} + 2q_{12} - q_{13})c_1 + 2\frac{q_{14}c_1^2 + q_{15}c_1c_3}{m_{\psi}^2}\Bigg]\,,\quad
s_7 = \frac{M^2}{16\pi^2}q_{14}\left(c_2 - \frac{c_1c_{\psi}}{m_{\psi}}\right)^2\,,\nonumber\\
s_1 = q_{16}f_1 + \frac{M^2c_{\psi}}{16\pi^2m_{\psi}^2}\left[(q_9m_{\psi}^2 + q_{15})(c_4m_{\psi}^2 - c_3c_{\psi}) + q_{12}(c_2m_{\psi}^2 - c_1c_{\psi})m_{\psi}^2\right]\,,\label{eqn:h4002coeff}\\
s_3 = s_5m_{\psi} = \frac{M^2c_{\psi}}{32\pi^2m_{\psi}^3}\left[q_7((c_3 - c_1)c_{\psi} + (c_2 - c_4)m_{\psi}^2)m_{\psi}^2 - 2(q_{11}m_{\psi}^2 + 2q_{14}c_1)(c_1c_{\psi} - c_2m_{\psi}^2)\right]\,,\nonumber\\
s_4 = 2s_6m_{\psi} = \frac{M^2c_{\psi}^2}{16\pi^2m_{\psi}^3}\left[2q_3m_{\psi}^4 + q_7(c_1 - c_3)m_{\psi}^2 + 2q_{11}c_1m_{\psi}^2 + 2q_{14}c_1^2\right]\,,\nonumber\\
s_8 = q_{16}f_3\,,\quad
s_9 = q_{16}f_4\,,\quad
s_{10} = q_{16}f_5\,,\quad
s_{11} = q_{16}f_6\,.\nonumber
\end{gather}
The solution~\eqref{eqn:h400sol} for \(h^{(4)}_{00}\) is given in terms of the following coefficients:
\begin{gather}
u_7 = -s_3 + s_5m_{\psi} - \frac{s_8 + s_{10}}{m_{\psi}^2}\,,\quad
u_8 = u_4m_{\psi} = s_1 - s_3m_{\psi} + \frac{s_5m_{\psi}^2}{2} + \frac{s_8 - s_{10}}{m_{\psi}}\,,\nonumber\\
u_{10} = -s_4 + 2s_6m_{\psi} - \frac{s_9 + s_{11}}{m_{\psi}^2}\,,\quad
u_{11} = 2u_5m_{\psi} = s_2 - 2s_4m_{\psi} + 2s_6m_{\psi}^2 + \frac{s_9 - s_{11}}{m_{\psi}}\,,\label{eqn:h400solcoeff}\\
u_1 = \frac{s_7}{2}\,,\quad
u_2 = \frac{s_5}{2}\,,\quad
u_3 = \frac{s_6}{2}\,,\quad
u_6 = \frac{s_8}{m_{\psi}^2}\,,\quad
u_9 = \frac{s_9}{m_{\psi}^2}\,,\quad
u_{12} = \frac{s_{10}}{m_{\psi}^2}\,,\quad
u_{13} = \frac{s_{11}}{m_{\psi}^2}\,.\nonumber
\end{gather}
In the case of a massless scalar field one obtains equation~\eqref{eqn:h400ii} with the following coefficients:
\begin{gather}
w_3 = \frac{2q_{19}c_2 - 2q_4c_2^2 - q_8c_{\psi}\left(c_2 - 3c_4 + 2c_{\psi}\frac{G_{4(1,0)}}{G_{4(0,0)}}\right) - 2q_9c_4c_{\psi} - (2q_{12} - q_{13})c_2c_{\psi} - 2q_{15}c_2c_4}{8\pi}\,,\nonumber\\
w_2 = \frac{2q_3c_{\psi}^2 + q_7(c_2 - c_4)c_{\psi} + 2q_{11}c_2c_{\psi} + 2q_{14}c_2^2}{32\pi^2}\,,\label{eqn:h400iicoeff}\\
w_1 = \frac{q_1c_{\psi}}{4\pi}\,,\quad
w_4 = q_{17}\,,\quad
w_5 = q_{18}\,,\quad
w_6 = q_{20}\,.\nonumber
\end{gather}

\acknowledgments
The author is happy to thank Xian Gao, Laur J\"arv and Ott Vilson for valuable feedback. He gratefully acknowledges the full financial support of the Estonian Research Council through the Postdoctoral Research Grant ERMOS115 and the Startup Research Grant PUT790.

\end{document}